\documentclass[a4,12pt]{article}

\usepackage{a4}

\usepackage{amsmath}
\usepackage{amsfonts}
\usepackage{amssymb}

\def\tr{\mathop{\rm tr}\nolimits}

\newcommand{\CC}{\mathbb{C}}
\newcommand{\ZZ}{\mathbb{Z}}
\newcommand{\RR}{\mathbb{R}}
\newcommand{\PP}{\mathbb{P}}

\usepackage{bm}

\newcommand{\wt}{\widetilde}

\def\Pexp{\mathop{\rm Pexp}\nolimits}

\begin{document}
\begin{titlepage}
\title{
\vspace{-1.5cm}
\begin{flushright}
{\normalsize TIT/HEP-681\\ July 2020}
\end{flushright}
\vspace{1.5cm}
\LARGE{Finite-$N$ corrections to the M-brane indices}}
\author{
Reona {\scshape Arai$^1$\footnote{E-mail: r.arai@th.phys.titech.ac.jp}},
Shota {\scshape Fujiwara$^1$\footnote{E-mail: s.fujiwara@th.phys.titech.ac.jp}},
Yosuke {\scshape Imamura$^1$\footnote{E-mail: imamura@phys.titech.ac.jp}},\\
Tatsuya {\scshape Mori$^1$\footnote{E-mail: t.mori@th.phys.titech.ac.jp}},
and
Daisuke {\scshape Yokoyama$^2$\footnote{E-mail: ddyokoyama@meiji.ac.jp}}
\\
\\
{\itshape $^1$Department of Physics, Tokyo Institute of Technology,} \\ {\itshape Tokyo 152-8551, Japan} \\
\textit{and} \\ \textit{ $^2$Department of Physics, Meiji University,} \\ \textit{ Kanagawa 214-8571, Japan}}

\date{}
\maketitle
\thispagestyle{empty}

\begin{abstract}
We investigate finite-$N$ corrections to the superconformal indices of the theories
realized on M2- and M5-branes.
For three-dimensional theories realized on a stack of $N$ M2-branes we calculate
the finite-$N$ corrections as the contribution of extended M5-branes in
the dual geometry $AdS_4\times \bm{S}^7$.
We take only M5-brane configurations with a single wrapping into account,
and neglect multiple-wrapping configurations.
We compare the results with the indices calculated from the ABJM theory,
and find agreement up to expected errors due to the multiple wrapping.
For six-dimensional theories on $N$ M5-branes
we calculate the indices by analyzing extended M2-branes in
$AdS_7\times \bm{S}^4$.
Again, we include only configurations with single wrapping.
We first compare the result for $N=1$ with the index of the free tensor multiplet
to estimate the order of the error due to multiple wrapping.
We calculate first few terms of the index of $A_{N-1}$ theories explicitly,
and confirm that they can be expanded by superconformal representations.
We also discuss multiple-wrapping contributions to the six-dimensional
Schur-like index.
\end{abstract}
\end{titlepage}
\tableofcontents
\section{Introduction}
In typical utilization of the AdS/CFT correspondence \cite{Maldacena:1997re}
we calculate quantities in the boundary theory
by using the gravity or string theory in the bulk.
For this to be possible it is necessary
that the quantum gravitational effect
is suppressed
because we do not have enough knowledge to carry out
quantitative analysis of quantum gravity.
Due to this restriction
the majority of works about the AdS/CFT correspondence
assume the large-$N$ limit.

However, there is a possibility that
some physical quantities in supersymmetric theories
are protected from the quantum gravity corrections
and we can perform an analysis on the gravity side
even if $N$ is finite.
An example of such a quantity is the BPS partition function
of the four-dimensional ${\cal N}=4$ supersymmetric Yang-Mills theory.
It was shown in \cite{Biswas:2006tj} that by geometric quantization
of BPS configurations of D3-branes expanded in $\bm{S}^5$
we can reproduce the exact BPS partition function for finite $N$.

Based on the result of \cite{Biswas:2006tj} two of the authors
proposed a prescription to calculate the finite-$N$
corrections to the superconformal index \cite{Kinney:2005ej} of ${\cal N}=4$ SYM
as the contribution of D3-branes wrapped on topologically trivial cycles
in $\bm{S}^5$ \cite{Arai:2019xmp}.
The method was also applied to S-fold theories
\cite{Garcia-Etxebarria:2015wns} and
the consistency with the
supersymmetry enhancement \cite{Aharony:2016kai}
was confirmed.
Later, it was extended to orbifold theories \cite{Arai:2019wgv}
and toric gauge theories \cite{Arai:2019aou}, and it was found that the
prescription works at least for
single-brane configurations.
The contribution of multiple-brane configurations was
first calculated in \cite{Arai:2020qaj} in the Schur limit,
and the analytic result in \cite{Bourdier:2015wda}
\begin{align}
\left.\frac{{\cal I}_N}{{\cal I}_\infty}\right|_{\rm Schur}\hspace{-1em}
=1+\sum_{n=1}^\infty c_nq^{nN+n^2},\quad
c_n=(-1)^n\frac{N+2n}{n!}\prod_{k=1}^{n-1}(N+k),
\label{bcf}
\end{align}
was successfully
reproduced.

For analysis of D3-brane configurations in $AdS_5\times \bm{S}^5$
there are three important parameters:
the $AdS_5$ radius $L$, the $\bm{S}^5$ radius $r$, and the D3-brane tension $T_{\rm D3}$.
Two dimensionless parameters defined with them are
\begin{align}
\frac{L}{r}=1,\quad
V_3r^4T_{\rm D3}=N,
\label{d3brane}
\end{align}
where $V_3=2\pi^2$ is the volume of the unit $3$-sphere\footnote{%
The volume of the unit $n$-sphere $V_n$ is given for small $n$ by
$V_2=4\pi$, $V_3=2\pi^2$, $V_4=\frac{8\pi^2}{3}$, $V_5=\pi^3$, $V_6=\frac{16\pi^3}{15}$, and
$V_7=\frac{\pi^4}{3}$.}.
The ratio $L/r$ is the unit of the energy of
Kaluza-Klein gravitons in $\bm{S}^5$ normalized by the AdS radius,
and on the boundary theory point of view it is
interpreted as the scale dimension of a free scalar field $\phi$,
which is $1$ in the four-dimensional theory.
The second equation shows that the energy of a D3-brane wrapped around
a large $\bm{S}^3\subset \bm{S}^5$ is as $N$ times as that of a unit Kaluza-Klein mode,
and we identify the wrapped brane with an operator like $\det \phi$.
In the large-$N$ limit such wrapped D3-branes
decouple while
for finite $N$
they are expected to contribute to the superconformal index
as finite-$N$ corrections.

In the prescription proposed in \cite{Arai:2019xmp,Arai:2020qaj}
the complete superconformal index is given by
\begin{align}
{\cal I}={\cal I}_{\rm KK}\left(1+\sum_C{\cal I}^{\rm D3}_C\right).
\label{id3}
\end{align}
${\cal I}_{\rm KK}$ is the index of the supergravity Kaluza-Klein modes,
which reproduces the gauge theory index in the large-$N$ limit \cite{Kinney:2005ej}.
$C$ runs over a set of D3-brane configurations.
Each configuration consists of D3-branes wrapped around large $\bm{S}^3$ in $\bm{S}^5$.
${\cal I}_C^{\rm D3}$ consists of two factors;
${\cal I}_C^{\rm D3}={\cal I}_C^{\rm ground}{\cal I}_C^{\rm excitations}$.
The factor ${\cal I}_C^{\rm ground}$ gives the
classical contribution from wrapped D3-branes without fluctuations.
If $C$ consists of $n$ wrapped D3-branes it is proportional to $q^{nN}$,
where $q$ is the fugacity associated with the energy.
The other factor ${\cal I}_C^{\rm excitations}$ is the index of the theory realized on
the configuration $C$.
If $C$ consists of a single D3-brane the theory is free $U(1)$ supersymmetric
gauge theory and the index is given by
\begin{align}
{\cal I}_C^{\rm excitations}=\Pexp i_C^{\rm D3},
\end{align}
where $i_C^{\rm D3}$ is the single-particle index of the fluctuation modes on the brane.
$\Pexp$ is the plethystic exponential defined as follows.
Let $f$ be a function of fugacities given as a formal power series
\begin{align}
f=\sum_nc_nx_n,
\end{align}
where $x_n$ are products of fugacities and
$c_n$ are integer coefficients.
Then, the plethystic exponential of $f$ is defined by
\begin{align}
\Pexp f=\prod_n(1-x_n)^{-c_n}.
\label{qexpdef}
\end{align}

Because a large $\bm{S}^3$ is a topologically trivial
cycle in $\bm{S}^5$
and a D3-brane wrapped around such a cycle is shrinkable
there exist tachyonic modes with negative energy on $C$ consisting of such branes.
The treatment of such tachyonic modes is a key point to
calculate the finite-$N$ corrections.
A simple analysis shows that on a single wrapped D3-brane
there is one tachyonic mode and its energy is $-1$ in the unit of $L^{-1}$.
Correspondingly, the single-particle index $i$ includes
the term $q^{-1}$. (We set other fugacities to be $1$.)
With the definition (\ref{qexpdef})
the plethystic exponential of this negative-power
term is
\begin{align}
\Pexp(q^{-1})=\frac{1}{1-q^{-1}}=-\frac{q}{1-q}.
\label{expt}
\end{align}
Interestingly, this factor increases
the order of the index by $1$ and changes the overall sign of the correction.
Although these facts are against intuition
the corrections calculated in this way agree with
known results.

An interpretation of the correction is as follows.
In the large-$N$ limit the complete index is reproduced by
the Kaluza-Klein index ${\cal I}_{\rm KK}$.
If $N$ is finite we should consider giant gravitons \cite{McGreevy:2000cw} instead of
the supergravity Kaluza-Klein modes.
An important difference from the supergravity index is the existence of
the upper bound of the size of giant gravitons.
Namely, the finite-$N$ index is obtained by somehow
subtracting
contributions of high momentum modes that do not
have corresponding giant gravitons.
The negative correction including the factor (\ref{expt}) is interpreted as the absence
of giant gravitons of large momenta.

If there are $\delta$ tachyonic modes
they raise the order of the correction
by $\delta$.
The interesting exponent $nN+n^2$ in (\ref{bcf}) can be
interpreted as the effect of $\delta=n^2$ tachyonic modes
on the configuration with $n$ D3-branes \cite{Arai:2020qaj}.
In the following we
call the shift $\delta$ of the order of a correction
``the tachyonic shift.''

The purpose of this paper is to apply the same idea
to the theories on M2-branes and M5-branes.
The BPS partition functions of these theories were
calculated in \cite{Bhattacharyya:2007sa},
and were reproduced by the geometric quantization of
M5- and M2-branes, respectively, in the same reference.
We extend their analysis to the superconformal indices
according to the prescription in \cite{Arai:2019xmp}.

This paper is organized as follows.
In Section \ref{abjm.sec}, we investigate the finite-$N$ corrections of
the M2-brane theories.
We first derive the formula for the finite-$N$ corrections induced by
a single wrapped M5-brane.
We compare the results obtained by the formula
with the index of ABJM theory with $k=1$ \cite{Aharony:2008ug},
and find nice agreement.
We also consider $\ZZ_k$ orbifolds corresponding to the
ABJM theory with the Chern-Simons level $k=2$ and $3$.
The comparison of the gravitational analysis and localization formula
again find nice agreement.

In Section \ref{an6d.sec}, we consider the finite-$N$ corrections of
the 6d ${\cal N}=(2,0)$ theories.
We first derive the formula for the finite-$N$ corrections induced by
a single wrapped M2-brane, and estimate the error due to the multiple wrapping by using $N=1$ case.
Then, we calculate the index of $A_{N-1}$ theories
by using the formula.
As a consistency check we confirm that they are expanded by
indices of superconformal irreducible representations.
We also consider the Schur-like limit of the index and discuss multiple wrapping contributions.

In Section \ref{disc.sec}, we summarize the results and discuss some extensions.

In Appendices we show some technical detailks and results that we do not show in the main text.

\section{3d ${\cal N}=8$ superconformal theories}\label{abjm.sec}

The three-dimensional ${\cal N}=8$ superconformal
theory realized
on a stack of $N$ coincident M2-branes
is described by the ABJM theory with Chern-Simons level $k=1$ \cite{Aharony:2008ug}.
The gravity dual is the M-theory in the $AdS_4\times \bm{S}^7$ background.
The $AdS_4$ radius $\hat L$, the $\bm{S}^7$ radius $\hat r$,
and the M5-brane tension $T_{\rm M5}$ satisfy the
relations
\begin{align}
\frac{\hat L}{\hat r}=\frac{1}{2},\quad
V_5\hat r^6T_{\rm M5}=N.
\label{ads4params}
\end{align}
(We use hats for distinction from similar
symbols used in the next section,
in which we will use checked symbols.)
With these relations we can easily see that the energy of a maximum giant M5-brane has energy $N/2$.
This fact suggests that such wrapped M5-branes correspond to
baryonic type operators in the ABJM theory \cite{Aharony:2008ug}.

In this section we investigate the fluctuations on such a wrapped M5-brane
and calculate the finite-$N$ corrections to the superconformal index.
We compare the results with the index
obtained by using localization formula.
We will focus on configurations
consisting of a single wrapped M5-brane, and
we will not consider
configurations with multiple wrapped M5-branes.

\subsection{Superconformal index}
The 3d ${\cal N}=8$ superconformal algebra
is $\hat{\cal A}=osp(8|4)$,
whose bosonic subalgebra is
$so(2,3)\times so(8)\subset\hat{\cal A}$.
There are six Cartan generators
\begin{align}
\hat H,\quad
\hat J_{12},\quad
\hat R_{12},\quad
\hat R_{34},\quad
\hat R_{56},\quad
\hat R_{78}.
\end{align}
The Hamiltonian $\hat H$ and the spin $\hat J_{12}$ are Cartan generator of $so(2,3)$ and
the other four are Cartan generators of the R-symmetry $so(8)$.
To define the superconformal index we choose one
complex supercharge $\hat {\cal Q}$ that carries specific Cartan charges.
We take the one with the following quantum numbers:
\begin{align}
\hat{\cal Q}:(\hat H,\hat J_{12};\hat R_{12},\hat R_{34},\hat R_{56},\hat R_{78})
=(
+\tfrac{1}{2},
-\tfrac{1}{2};
+\tfrac{1}{2},
+\tfrac{1}{2},
+\tfrac{1}{2},
+\tfrac{1}{2}).
\label{qhat}
\end{align}
The subalgebra of $\hat{\cal A}$
that keeps the chosen supercharge $\hat{\cal Q}$ intact is
\begin{align}
\hat{\cal B}\times u(1)_{\hat\Delta}\subset \hat{\cal A},
\end{align}
where $\hat{\cal B}=osp(6|2)$ is the superalgebra whose bosonic subalgebra is $sl(2,\RR)\times so(6)$.
The central factor $u(1)_{\hat\Delta}$
is generated by
\begin{align}
\hat\Delta=\{\hat{\cal Q},\hat{\cal Q}^\dagger\}
=\hat H-\hat J_{12}-\frac{1}{2}(\hat R_{12}+\hat R_{34}+\hat R_{56}+\hat R_{78}).
\end{align}
The superconformal index associated with
the BPS bound $\hat\Delta\geq0$ is defined as the $\hat{\cal B}$ character by%
\footnote{The fugacities used here are related to those in
Section 2 of \cite{Bhattacharya:2008zy} by
$\hat q=x$,
$\hat u_1=y_1^{-\frac{1}{2}}y_2^{\frac{1}{2}}y_3^{\frac{1}{2}}$,
$\hat u_2=y_1^{\frac{1}{2}}y_2^{-\frac{1}{2}}y_3^{\frac{1}{2}}$,
$\hat u_3=y_1^{\frac{1}{2}}y_2^{\frac{1}{2}}y_3^{-\frac{1}{2}}$,
and
$\hat u_4=y_1^{-\frac{1}{2}}y_2^{-\frac{1}{2}}y_3^{-\frac{1}{2}}$.}
\begin{align}
{\cal I}(\hat q,\hat u_i)=\tr[(-1)^F\hat x^{\hat\Delta}\hat q^{\hat H+\hat J_{12}}
\hat u_1^{\hat R_{12}}
\hat u_2^{\hat R_{34}}
\hat u_3^{\hat R_{56}}
\hat u_4^{\hat R_{78}}
],\quad
\hat u_1\hat u_2\hat u_3\hat u_4=1.
\label{sci3d}
\end{align}
Due to the Bose-Fermi degeneracy for $\hat\Delta>0$
this does not depend on $\hat x$.

\subsection{Wrapped M5-branes}
In the large-$N$ limit
the superconformal index is reproduced by
the Kaluza-Klein modes in $AdS_4\times \bm{S}^7$.
The Kaluza-Klein index ${\cal I}_{\rm KK}$
is given by ${\cal I}_{\rm KK}=\Pexp i_{\rm KK}$, where $i_{\rm KK}$ is
the single-particle index \cite{Bhattacharya:2008zy}
\begin{align}
i_{\rm KK}=
\frac{(1-\hat q^{\frac{3}{2}}\hat u_1^{-1})(1-\hat q^{\frac{3}{2}}\hat u_2^{-1})(1-\hat q^{\frac{3}{2}}\hat u_3^{-1})(1-\hat q^{\frac{3}{2}}\hat u_4^{-1})}
{(1-\hat q^{\frac{1}{2}}\hat u_1)(1-\hat q^{\frac{1}{2}}\hat u_2)(1-\hat q^{\frac{1}{2}}\hat u_3)(1-\hat q^{\frac{1}{2}}\hat u_4)(1-\hat q^2)^2}
-\frac{1-\hat q^2+\hat q^4}{(1-\hat q^2)^2}.
\end{align}
The corresponding boundary theory is
the ABJM theory with the Chern-Simons level $k=1$.
The full index of the ABJM theory including
the contribution of monopole operators
was calculated in \cite{Kim:2009wb},
and the agreement of the ABJM index in the large-$N$ limit
${\cal I}^{\rm ABJM}_{N=\infty}$ and this Kaluza-Klein
index ${\cal I}_{\rm KK}$ was confirmed.

Based on the idea in \cite{Arai:2019xmp}
we propose the following equation
for the finite-$N$ index
\begin{align}
{\cal I}_N^{\rm ABJM}={\cal I}_{\rm KK}
\left(1+\sum_C{\cal I}^{\rm M5}_C\right).
\label{m5correction}
\end{align}
The second term in the parentheses
in (\ref{m5correction}) gives
the finite-$N$ corrections due to wrapped M5-branes.
$C$ runs over ``the representative configurations''
of wrapped M5-branes
specified in the following, and ${\cal I}_C^{\rm M5}$
is the contribution of each configuration $C$.

We determine the representative configurations $C$
by a preliminary analysis of a rigid M5-brane,
an M5-brane wrapped on a large $\bm{S}^5$ in $\bm{S}^7$.
Let us introduce complex coordinates $z_a$ ($a=1,2,3,4$)
to describe the $\bm{S}^7$ by
$\sum_{a=1}^4|z_a|^2=1$.
The R-symmetry $su(4)\subset\hat{\cal B}$ acts on these coordinates
in the natural way.
For a rigid M5-brane to
be BPS with respect to the chosen supercharge $\hat{\cal Q}$
the worldvolume must be given by the holomorphic
equation \cite{Mikhailov:2000ya}
\begin{align}
c_1z_1+c_2z_2+c_3z_3+c_4z_4=0,
\end{align}
were $c_a$ are homogeneous coordinates in $\PP^3$.
The collective motion of the M5-brane can be treated
as a particle in the moduli space $\PP^3$.
By the analysis of the coupling of the brane and the
background flux we find
the wave function $\Psi$ of a rigid M5-brane
is a section of the line bundle ${\cal O}(N)$ over $\PP^3$.
We can give $\Psi$ as a homogeneous function of
the coordinates $c_a$ of degree $N$.
States described by such wave functions belong to
the $su(4)$ representation with Dynkin labels $[N,0,0]$.
On the gauge theory side these states
are identified with baryonic type operators in the ABJM theory\cite{Aharony:2008ug}.
The corresponding index is
$\hat q^{\frac{1}{2}N}\chi_{[N,0,0]}(\hat u_a)$,
where $\chi_{[a,b,c]}(\hat u_a)$ is the $su(4)$ character of the representation
$[a,b,c]$.
The characters of the fundamental
and the anti-fundamental representations are given by
\begin{align}
\chi_{[1,0,0]}(\hat u)=
\hat u_1+\hat u_2+\hat u_3+\hat u_4,\quad
\chi_{[0,0,1]}(\hat u)=
\hat u_1^{-1}+\hat u_2^{-1}+\hat u_3^{-1}+\hat u_4^{-1}.
\label{fafcharacters}
\end{align}

Now let us remember the Weyl's character formula.
It gives $\hat{q}^{\frac{1}{2}N}\chi_{[N,0,0]}(\hat{u}_a)$
as the sum:
\begin{align}
\hat q^{\frac{1}{2}N}\chi_{[N,0,0]}(\hat u_a)
&=\frac{\hat q^{\frac{1}{2}N}\hat u_4^N}{(1-\frac{\hat u_1}{\hat u_4})(1-\frac{\hat u_2}{\hat u_4})(1-\frac{\hat u_3}{\hat u_4})} +(\mbox{permutations})\nonumber\\
&=\hat q^{\frac{1}{2}N}\hat u_4^N\Pexp\left(\frac{\hat u_1}{\hat u_4}+\frac{\hat u_2}{\hat u_4}+\frac{\hat u_3}{\hat u_4}\right) +(\mbox{permutations}),
\label{su4weyl}
\end{align}
where ``permutations'' represents three terms obtained
from the first term by cyclic permutations of $\hat u_a$.
From the quantum mechanical point of view,
the first term can be interpreted as the partition
function of the system with
the ground state $\hat q^{\frac{1}{2}N}\hat u_4^N$ and three bosonic excitations $\hat u_1/\hat u_4$, $\hat u_2/\hat u_4$, and $\hat u_3/\hat u_4$.
We define the representative configuration as the M5-brane corresponding to the ground state.
For the first term in (\ref{su4weyl}) it is given by $z_4=0$.
Corresponding to the other terms obtained by the permutations
there are three more representative configurations $z_a=0$ ($a=1,2,3$).

The main idea in \cite{Arai:2019xmp} is that we can obtain the finite-$N$ corrections
to the index by ornamenting the Weyl's formula (\ref{su4weyl}) with
all other fluctuation modes
by replacing the zero-mode contribution
$\hat u_1/\hat u_4+\hat u_2/\hat u_4+\hat u_3/\hat u_4$ by the complete single-particle index
of the theory on the worldvolume of the M5-brane.
In addition, to obtain the complete corrections, we need to
take account of representative configurations including more than one branes \cite{Arai:2020qaj}.
Namely, the general form of $C$ is given by
\begin{align}
C:
z_1^{n_1}
z_2^{n_2}
z_3^{n_3}
z_4^{n_4}=0,
\quad n_a\in\ZZ_{\geq0},\quad
(n_1,n_2,n_3,n_4)\neq(0,0,0,0),
\label{repconfig}
\end{align}
where a multiple zero is understood as coincident branes.
$(n_1,n_2,n_3,n_4)=(0,0,0,0)$ is excluded because it corresponds to
the first term in the parentheses in (\ref{m5correction}).
The contribution of each configuration $C$ is
factorized into two factors ${\cal I}_C^{\rm ground}$ and
${\cal I}_C^{\rm excitations}$.
Each wrapped brane contributes $N/2$ to the energy (in the unit of $\hat L^{-1}$)
and the ground state of $C$ includes the factor $\hat q^{\frac{1}{2}nN}$
with $n=n_1+n_2+n_3+n_4$.
${\cal I}_C^{\rm ground}$
is given as the product of the
ground state contribution of each brane:
\begin{align}
{\cal I}_C^{\rm ground}
=\hat q^{\frac{1}{2}nN}
\hat u_1^{n_1N}
\hat u_2^{n_2N}
\hat u_3^{n_3N}
\hat u_4^{n_4N}.
\end{align}
${\cal I}_C^{\rm excitations}$
is the contribution of excitations
on $C$.
If $n\geq2$ the theory on $C$ is interacting
and it is not so easy to
calculate ${\cal I}_C^{\rm excitations}$.
In this work we only consider four configurations with $n=1$ given by $z_a=0$ ($a=1,2,3,4$).
Then, the theory on $C$ is free and
${\cal I}_C^{\rm excitations}$ is given by
\begin{align}
{\cal I}_{z_a=0}^{\rm excitations}=\Pexp i_{z_a=0}^{\rm M5},
\end{align}
where $i^{\rm M5}_{z_a=0}$ is the single-particle index
of the fluctuation modes on the worldvolume of an M5-brane wrapped on $z_a=0$.

Let us calculate the single-particle index
$i_{z_a=0}^{\rm M5}$ for each representative configuration.
In the following we consider the configuration $z_4=0$.
The other three are obtained by the permutations of the fugacities $\hat u_a$.
We start with the analysis of the scalar modes.
If we neglect the self-dual potential field and fermion fields on the worldvolume
the M5-brane action is given as the sum of the Nambu-Goto action $S_{\rm NG}$ and
the Chern-Simons term $S_{\rm CS}$:
\begin{align}
S_{\rm NG}=-T_{\rm M5}\int d^6\sigma\sqrt{-\det G_{ab}},\quad
S_{\rm CS}=\int A_6,
\end{align}
where $G_{ab}$ is the induced metric and $A_6$ is the background $6$-form
potential satisfying $dA_6=(2\pi N/V_7){\rm vol}(\bm{S}^7)$.
We use the following  AdS$_4\times \bm{S}^7$ metric:
\begin{align}
ds^2=\hat{L}^2(-\cosh^2\rho d\hat{t}^2+d\rho^2+\sinh^2\rho d\Omega_2^2)
+\hat{r}^2(\cos^2\theta d\Omega_5^2+d\theta^2+\sin^2\theta d\phi^2).
\label{ads4s7metric}
\end{align}

We consider an M5-brane wrapped on $\RR\times \bm{S}^5$ defined by $\rho=\theta=0$.
There are $5$ scalar fields corresponding to transverse directions of the M5-brane:
three in $AdS_4$ and two in $\bm{S}^7$.
To describe fluctuations in $AdS_4$ we
introduce a three-dimensional unit vector $\bm{n}$ and rewrite
$d\Omega_2^2$ as $d\bm{n}^2$.
We define fluctuation fields by
\begin{align}
\bm{X}=\rho\bm{n},\quad
z=\theta e^{i\phi}.
\end{align}
By neglecting higher order terms and
using the relations in
(\ref{ads4params})
 we obtain
\begin{align}
S_{\rm NG}
+S_{\rm CS}
=&\frac{N}{2\pi^3}
\int d\hat{t}d\Omega_5
\Bigg[-1
+\frac{1}{2}(\partial_{\hat{t}}\bm{X})^2
-\frac{1}{8}(\nabla\bm{X})^2
-\frac{1}{2}\bm{X}^2
\nonumber\\&\hspace{3em}
+2|\partial_{\hat{t}}z|^2
-\frac{1}{2}|\nabla z|^2
+\frac{5}{2}|z|^2
+3i(-z^*\partial_{\hat{t}}z+z \partial_{\hat{t}}z^*)
\Bigg],
\label{quadraticm5}
\end{align}
where $\nabla$ is the derivative on the unit $\bm{S}^5$.
The constant term gives the energy $E=\frac{1}{2}N$ of the wrapped M5-brane.
By solving the equations of motion we can easily obtain the spectrum of
fluctuation modes.
(See Table \ref{m5modes}.)
\begin{table}[htb]
\caption{Scalar fluctuation modes on an M5-brane wrapped on $z_4=0$.
$\ell=0,1,2,\ldots$ is the angular momentum on $\bm{S}^5$.}\label{m5modes}
\centering
\begin{tabular}{ccccc}
\hline
\hline
fields & $\hat J_{12}$ & $so(6)$ & $\hat R_{78}$ & $\hat H$ \\
\hline
$\bm{X}$ & $0,\pm 1$ & $[0,\ell,0]$ & $0$ & $(\ell+2)/2$ \\
$z$   & $0$ & $[0,\ell,0]$ & $+1$ & $(\ell+5)/2$ \\
$z^*$ & $0$ & $[0,\ell,0]$ & $-1$ & $(\ell-1)/2$ \\
\hline
\end{tabular}
\end{table}
We have six zero-modes of $z^*$ at $\ell=1$
and three of them are BPS.
They correspond to three excitations
$\hat{u}_1/\hat u_4$,
$\hat{u}_2/\hat u_4$, and
$\hat{u}_3/\hat u_4$ appearing in the Weyl's formula (\ref{su4weyl}).
We also have one BPS tachyonic mode of $z^*$ at $\ell=0$.

A few comments on the tachyonic mode are in order.
First, the existence of the tachyonic mode
does not cause the instability of the system.
The tachyonic mode carries the R-charge $\hat R_{78}=-1$,
and a tachyonic particle is always created together with an
anti-particle with $\hat R_{78}=+1$.
As is shown in Table \ref{m5modes} such an anti-larticle,
which corresponds to the $\ell=0$ mode of $z$, carries
the energy $E=5/2$, and the pair creation raises the total energy of the
system.
Another comment is about the consistency with the BPS bound.
Ordinarily, a particle with negative energy is against the BPS bound $E\geq0$.
In the theory on the wrapped brane, however, we do not have such a bound.
An M5-brane wrapped on $z_4=0$
breaks the half supersymmetries.
Among $32$ supercharges only $16$ that commute
with the generator
\begin{align}
\hat Z=\hat H-\hat R_{78}
\end{align}
are preserved.
The algebra of the preserved symmetry is
\begin{align}
\hat{\cal C}\times u(1)_{\hat Z},\quad
\hat{\cal C}=su(2|4).
\label{m5unbroken}
\end{align}
The central factor $u(1)_{\hat Z}$ is generated
by $\hat Z$.
The bosonic subalgebra of $\hat{\cal C}$
is $so(3)\times so(6)\times u(1)$
generated by
\begin{align}
\hat J_{ij}\quad
(i,j=1,2,3),\quad
\hat R_{ab}\quad
(a,b=1,\ldots,6),\quad
\hat C\equiv \hat H-\frac{1}{2}\hat R_{78}.
\end{align}
The fluctuation modes on the M5-brane form
a representation of the unbroken algebra $\hat{\cal C}$.
The Hamiltonian $\hat H$ appears in $\hat{\cal C}$
only through $\hat C$, and the bound obtained from the
algebra is not $\hat H\geq0$ but $\hat C\geq0$.
The tachyonic mode satutates this bound.

In principle,
we can calculate the
complete single-particle index $i_{z_a=0}^{\rm M5}$
by carrying out the mode expansion
of the tensor and the fermion fields.
However, there is an easy way to obtain the index
from the known 6d superconformal index of the tensor multiplet.

We are interested in the
theory of a tensor multiplet living on
$\RR\times\bm{S}^5$, the worldvolume of a wrapped M5-brane.
This system is similar to the system of a tensor multiplet
living on the boundary of $AdS_7$.
In the next section we investigate the six-dimensional
system living on the AdS boundary $\RR\times\bm{S}^5$,
on which the
$(2,0)$ superconformal algebra $\check{\cal A}$
acts.
The two free theories, the theory on a wrapped M5-brane in $AdS_4\times \bm{S}^7$ and
the theory on the boundary of $AdS_7$, are in fact the same theory,
at least at the linearized level, and we can obtain the index of
the former from the index of the latter by a simple variable change of
fugacities.

We first establish the relation between the symmetry algebras.
Namely, we need to find an isomorphism between
the unbroken algebra on the wrapped M5-brane
(\ref{m5unbroken}) and a subalgebra of
$\check{\cal A}$.
There is an ambiguity of the choice of the subalgebra of $\check{A}$.
A convenient one is the symmetry (\ref{m2unbroken}) realized on
a wrapped M2-brane studied in the next section.
It is isomorphic to
(\ref{m5unbroken});
\begin{align}
\hat{\cal C}\times u(1)_{\hat Z}
\simeq
\check{\cal C}\times u(1)_{\check Z}.
\label{theisomorphism}
\end{align}
The explicit relations between the two sets of the bosonic generators
are as follows.
\begin{align}
&
\check J_{ij}=\hat R_{ij}\quad(i,j=1,\ldots,6),\quad
\check R_{a+2,b+2}=\hat J_{ab}\quad(a,b=1,2,3),\nonumber\\
&\check Z=2\hat Z,\quad
\check C=2\hat C.
\label{isomorphism}
\end{align}

We can relate two systems not only at the level of the symmetry but also at the
level of the Lagrangians.
The boundary metric of $AdS_7$ is
\begin{align}
ds^2\propto -d\check t^2+d\Omega_5^2.
\label{bdrmetric}
\end{align}
For distinction from $\hat t$ used in (\ref{quadraticm5})
we use $\check t$ for the time coordinate.
The Lagrangian of the five scalar fields $\phi_I$ ($I=1,\ldots,5$) living on this background
is
\begin{align}
{\cal L}\propto \sum_{I=1}^5\left[
(\partial_{\check t}\phi_I)^2
-(\nabla\phi_I)^2
-4\phi_I^2\right],
\label{lkag}
\end{align}
where the last term is the conformal coupling to the background curvature.
We simply relate the triplet fields by $\bm{X}\propto(\phi_3,\phi_4,\phi_5)$, while
in the relation between $z$ and $\phi_{1,2}$
we need to apply the time-dependent phase rotation
\begin{align}
z\propto e^{-3i\check t}(\phi_1+i\phi_2),
\end{align}
corresponding to the relation of two Hamiltonians $2\hat H=\check H-3\check R_{12}$
obtained from the last two equations in (\ref{isomorphism}).
In addition, we rescale the time coordinate by $\hat t=2\check t$
to match the background metric
(\ref{bdrmetric}) and
the metric on the wrapped M5-brane
\begin{align}
ds^2=\hat r^2\left(-\frac{1}{4}d\hat t^2+d\Omega_5^2\right)
\end{align}
obtained from (\ref{ads4s7metric}) by the restriction $\rho=\theta=0$.
Then,
we obtain the Lagrangian in (\ref{quadraticm5}) from (\ref{lkag}).

We can extend
the relations
(\ref{isomorphism})
to fermionic generators.
An important fact is that
the supercharges used to define the superconformal
indices on two sides are related by
\begin{align}
\check{\cal Q}=\sqrt2\hat{\cal Q}^\dagger,
\end{align}
and the relation
$\check\Delta
=2\hat\Delta$ immediately follows from this.
This implies that the superconformal
indices defined on two sides are essentially
the same.
Indeed, we can rewrite the six-dimensional index
(\ref{index6d}) to the three-dimensional index (\ref{sci3d})
by using the map
(\ref{isomorphism})
and the variable change
\begin{align}
\check q=\hat q^{\frac{3}{8}}\hat u_4^{-\frac{1}{4}},\quad
&\check y_1=\hat u_1\hat u_4^{\frac{1}{3}},\quad
\check y_2=\hat u_2\hat u_4^{\frac{1}{3}},\quad
\check y_3=\hat u_3\hat u_4^{\frac{1}{3}},\quad
\check u=\hat q^{-\frac{5}{4}}\hat u_4^{-\frac{1}{2}}.
\label{ads7toads4}
\end{align}
Applying the variable change
(\ref{ads7toads4})
to the index $i_{\rm bdr}^{\rm M5}$ in
(\ref{freetensorindex}) of the free tensor multiplet
we obtain the following single-particle
index for the excitations on
an M5-brane wrapped on $z_4=0$:
\begin{align}
i^{\rm M5}_{z_4=0}
&=
\frac
{
\hat q^{-\frac{1}{2}}\hat u_4^{-1}
-\hat q\hat u_4^{-1}(\hat u_1^{-1}+\hat u_2^{-1}+\hat u_3^{-1})
+\hat q^{\frac{3}{2}}\hat u_4^{-1}
+\hat q^2
}
{
(1-\hat q^{\frac{1}{2}}\hat u_1)
(1-\hat q^{\frac{1}{2}}\hat u_2)
(1-\hat q^{\frac{1}{2}}\hat u_3)
}
\nonumber\\
&=\frac{1}{\hat q^{\frac{1}{2}}\hat u_4}
+\frac{\hat u_1+\hat u_2+\hat u_3}{\hat u_4}
+\cdots.
\end{align}
The first few terms in the expansion correspond
to the tachyonic modes and rigid motion modes
obtained in the analysis of scalar fluctuations.

\subsection{Comparison with known results}
In the last subsection we obtained
the following hypothetical formula
\begin{align}
{\cal I}^{\rm ABJM}_N={\cal I}^{\rm grav}_N+{\cal O}(\hat q^{\frac{1}{2}(2N+\delta)}),
\label{hyp3d}
\end{align}
where the first term in the right-hand side is defined by
\begin{align}
{\cal I}^{\rm grav}_N
:={\cal I}_{\rm KK}
\left(1+\sum_{a=1}^4\hat q^{\frac{1}{2}N}\hat u_a^N\Pexp i_{z_a=0}^{\rm M5}\right),
\label{igravm2}
\end{align}
and the second term ${\cal O}(\hat q^{\frac{1}{2}(2N+\delta)})$
is the expected error due to the neglect of the
multiple-wrapping configurations with the tachyonic shift $\delta$.
Based on the experience in the D3-brane case
we expect $\delta$ is independent of $N$,
and this is directly confirmed below for small $N$.

Let us first give the results on the gauge theory side.
If $N=1$ we do not have to use the ABJM theory.
Instead, we can use free theory
of scalar fields and fermions living on an M2-brane.
The index is given by ${\cal I}_{N=1}^{\rm ABJM}=\Pexp i_{\rm bdr}^{\rm M2}$
with the single-particle index
\cite{Bhattacharya:2008zy}
\begin{align}
i_{\rm bdr}^{\rm M2}=
\frac{\hat q^{\frac{1}{2}}\chi_{[1,0,0]}(\hat u)-\hat q^{\frac{3}{2}}\chi_{[0,0,1]}(\hat u)}{1-\hat q^2}.
\label{n1m2index}
\end{align}

For $N\geq 2$ we need to use the ABJM theory with the Chern-Simons level $k=1$,
and sum up contributions of monopole operators
according to \cite{Kim:2009wb}.
See Appendix \ref{abjmformula.sec} for the explicit formula.
The results for $N=1,2,3$ are
\begin{align}
{\cal I}^{\rm ABJM}_{N=1}|_{\hat u=1}
&=1
+4\hat q^{\frac{1}{2}}
+10\hat q
+16\hat q^{\frac{3}{2}}
+19\hat q^2
+20\hat q^{\frac{5}{2}}
+26\hat q^3
+40\hat q^{\frac{7}{2}}
+49\hat q^4
\nonumber\\&
+{\cal O}(\hat q^{\frac{9}{2}}),\\
{\cal I}^{\rm ABJM}_{N=2}|_{\hat u=1}
&=1
+4\hat q^{\frac{1}{2}}
+20\hat q
+56\hat q^{\frac{3}{2}}
+139\hat q^2
+260\hat q^{\frac{5}{2}}
+436\hat q^3
+640\hat q^{\frac{7}{2}}
+954\hat q^4
\nonumber\\&
+1420\hat q^{\frac{9}{2}}
+2076\hat q^5
+{\cal O}(\hat q^{\frac{11}{2}}),\\
{\cal I}^{\rm ABJM}_{N=3}|_{\hat u=1}
&=1
+4\hat q^{\frac{1}{2}}
+20\hat q
+76\hat q^{\frac{3}{2}}
+239\hat q^2
+644\hat q^{\frac{5}{2}}
+1512\hat q^3
+3100\hat q^{\frac{7}{2}}
+5743\hat q^4
\nonumber\\&
+9856\hat q^{\frac{9}{2}}
+16182\hat q^5
+25988\hat q^{\frac{11}{2}}
+40764\hat q^6
+{\cal O}(\hat q^{\frac{13}{2}}).
\end{align}
In this section we only show the results with $\hat u_a=1$ for readability.
Refer to Appendix \ref{fullexp.sec} for the full expressions.

Let us first compare these results with the Kaluza-Klein index
\begin{align}
{\cal I}_{\rm KK}|_{\hat u=1}
&=1
+4\hat q^{\frac{1}{2}}
+20\hat q
+76\hat q^{\frac{3}{2}}
+274\hat q^2
+{\cal O}(\hat q^{\frac{5}{2}}).
\end{align}
We find the finite-$N$ corrections appear at $\hat q^{\frac{1}{2}(N+1)}$.
These are consistent with the contributions of a single wrapped brane with one tachyonic mode.
(\ref{igravm2}) gives
the following results for $N=1,2,3$.
\begin{align}
{\cal I}^{\rm grav}_{N=1}|_{\hat u=1}
&=1
+4\hat q^{\frac{1}{2}}
+10\hat q
+16\hat q^{\frac{3}{2}}
+19\hat q^2
+20\hat q^{\frac{5}{2}}
+26\hat q^3
+40\hat q^{\frac{7}{2}}
+5769\hat q^4
\nonumber\\&
+{\cal O}(\hat q^{\frac{9}{2}}),\\
{\cal I}^{\rm grav}_{N=2}|_{\hat u=1}
&=1
+4\hat q^{\frac{1}{2}}
+20\hat q
+56\hat q^{\frac{3}{2}}
+139\hat q^2
+260\hat q^{\frac{5}{2}}
+436\hat q^3
+640\hat q^{\frac{7}{2}}
+954\hat q^4
\nonumber\\&
+1420\hat q^{\frac{9}{2}}
+15518\hat q^5
+{\cal O}(\hat q^{\frac{11}{2}}),\\
{\cal I}^{\rm grav}_{N=3}|_{\hat u=1}
&=1
+4\hat q^{\frac{1}{2}}
+20\hat q
+76\hat q^{\frac{3}{2}}
+239\hat q^2
+644\hat q^{\frac{5}{2}}
+1512\hat q^3
+3100\hat q^{\frac{7}{2}}
+5743\hat q^4
\nonumber\\&
+9856\hat q^{\frac{9}{2}}
+16182\hat q^5
+25988\hat q^{\frac{11}{2}}
+70079\hat q^6
+{\cal O}(\hat q^{\frac{13}{2}}).
\end{align}
We find nice agreement.
The error appears at the order $\hat q^{\frac{1}{2}(2N+\delta)}$ with $\delta=6$.
As is expected $\delta$ is $N$-independent.
At present we have no explanation for this specific value of $\delta$.

\subsection{$\ZZ_k$ orbifold}\label{orbi.sec}
It is easy to extend our formula (\ref{igravm2}) to the orbifold $AdS_4\times \bm{S}^7/\ZZ_k$ with $k\geq 2$
defined by the orbifold action
\begin{align}
(z_1,z_2,z_3,z_4)\rightarrow
(\omega_kz_1,\omega_kz_2,\omega_k^{-1}z_3,\omega_k^{-1}z_4),\quad
\omega_k\equiv\exp\frac{2\pi i}{k}.
\label{zkorbi}
\end{align}
On the gauge theory side this is described by the ABJM theory with the Chern-Simons level $k\geq2$.

The Kaluza-Klein contribution ${\cal I}_{\rm KK}^{\ZZ_k}$ is given by
\begin{align}
{\cal I}_{\rm KK}^{\ZZ_k}=\Pexp{\cal P}_k i_{\rm KK},
\end{align}
where ${\cal P}_k$ is the $\ZZ_k$ projection operator defined for a function $g$ of $su(4)$ fugacities
$\hat u_a$ by
\begin{align}
{\cal P}_kg(\hat u_1,\hat u_2,\hat u_3,\hat u_4)
=\frac{1}{k}\sum_{i=0}^{k-1}g(\omega_k^i\hat u_1,\omega_k^i\hat u_2,\omega_k^{-i}\hat u_3,\omega_k^{-i}\hat u_4).
\end{align}
The representative configurations $C$ are given by (\ref{repconfig}),
and again we focus on the four single-wrapping configurations $z_a=0$ ($a=1,2,3,4$).
Due to the $\ZZ_k$ orbifolding, the worldvolume of the M5-brane becomes $\bm{S}^5/\ZZ_k$,
and the excitation is described by the projected single-particle index ${\cal P}_ki_{z_a=0}^{\rm M5}$.
Then, the projected index ${\cal I}^{\rm M5}_{z_a=0}$ is given by
\begin{align}
{\cal I}^{\rm M5}_{z_a=0}
=\hat q^{\frac{1}{2}N}\hat u_a^N\Pexp{\cal P}_k i^{\rm M5}_{z_a=0}.
\end{align}

Because of the non-trivial five-cycle homology $H_5(\bm{S}^7/\ZZ_k)=\ZZ_k$
we can classify states by the topological wrapping number $B\in\ZZ_k$ of M5-branes,
and we can calculate the index for each sector with specific $B$.
If a configuration $C$ is given by equation $h(z)=0$
the function $h(z)$ must have
a specific $\ZZ_k$ charge
for consistency with the $\ZZ_k$ orbifolding.
Namely, it must satisfy
\begin{align}
h(\omega_kz_1,\omega_kz_2,\omega_k^{-1}z_3,\omega_k^{-1}z_4)
=\omega_k^Bh(z_1,z_2,z_3,z_4)
\end{align}
with some $B\in\ZZ_k$.
Then, $B$ is the topological wrapping number of the worldvolume.
Among the four representative configurations with $n=1$,
$z_1=0$ and $z_2=0$ carry $B=+1$, and
$z_3=0$ and $z_4=0$ carry $B=-1$.

On the ABJM theory side $k$ is the Chern-Simons level,
and a wrapped M5-brane with $B\neq0$ corresponds to
a baryonic operator carrying $\ZZ_k$ baryonic charge $B$.
In the ABJM theory with the gauge group $U(N)_k\times U(N)_{-k}$
this baryonic symmetry is a part of gauge symmetry,
and baryonic operators are not gauge invariant.
In order to calculate the index with the contribution
of baryonic operators we need to use the
ABJM theory with the gauge group $(U(N)_k\times U(N)_{-k})/\ZZ_k$
where the $\ZZ_k$ quotient acts on the diagonal $U(1)$ symmetry
\cite{Tachikawa:2019dvq,Bergman:2020ifi}.
In the index calculation this quotient changes the quantization of
monopole charges.

The index of ABJM theory is calculated by summing up
contribution of different monopole charges \cite{Kim:2009wb}.
The monopole charges are labeled by
$2N$ GNO charges:
$(m_1.\ldots,m_N)$ for $U(N)_k$ and
$(\wt m_1,\ldots,\wt m_N)$ for $U(N)_{-k}$.
In the $U(N)_k\times U(N)_{-k}$ theory
all charges are integers,
while in the $(U(N)_k\times U(N)_{-k})/\ZZ_k$ theory
the quantization condition is given by
\begin{align}
m_\alpha,\wt m_\alpha\in \ZZ+\frac{B}{k},\quad
B\in\ZZ_k.
\end{align}
The index of the $B=0$ sector is the same as the index of
the $U(N)_k\times U(N)_{-k}$ ABJM theory, while
$B\neq0$ sector gives the index for baryonic operators,
which corresponds to the contribution of M5-branes with topological wrapping number $B$
on the gravity side.

In the following we calculate the indices for $k=2$ and $k=3$ on both sides of the
duality, and confirm the agreement up to the expected order of $\hat q$.
We use the notations ${\cal I}^{{\rm ABJM}(B/k)}_{N}$ and ${\cal I}^{{\rm grav}(B/k)}_N$
for the indices calculated on the two sides of the duality.

\subsubsection{$k=2$}
In the case of $k=2$ there are two sectors labeled by $B\in\ZZ_2$.

Let us first calculate the index of the $B=0$ sector.
The indices for $N=1,2,3$ are
\begin{align}
{\cal I}^{{\rm ABJM}(0/2)}_{N=1}|_{\hat u=1}
&=1
+ 10\hat q
+ 19\hat q^2
+{\cal O}(\hat q^3),
\\
{\cal I}^{{\rm ABJM}(0/2)}_{N=2}|_{\hat u=1}
&=1
+10\hat q
+75\hat q^2
+220\hat q^3
+{\cal O}(\hat q^4),
\\
{\cal I}^{{\rm ABJM}(0/2)}_{N=3}|_{\hat u=1}
&=1
+10\hat q
+75\hat q^2
+450\hat q^3
+1595\hat q^4
+{\cal O}(\hat q^5).
\end{align}
Let us compare these with the Kaluza-Klein contribution.
\begin{align}
{\cal I}_{\rm KK}^{\ZZ_2}|_{\hat u=1}
&=1
+10\hat q
+75\hat q^2
+450\hat q^3
+2365\hat q^4
+{\cal O}(\hat q^5).
\end{align}
We find the corrections appear at order $\hat q^{\frac{1}{2}(2N+2)}$.
They are interpreted as contributions of two-brane configurations,
which belong to the $B=0$ sector.
Hence, it exceeds our scope.

Next, let us consider the index of $B=1$ sector:
\begin{align}
{\cal I}^{{\rm ABJM}(1/2)}_{N=1}|_{\hat u=1}
&=4\hat q^{\frac{1}{2}}
+16\hat q^{\frac{3}{2}}
+20\hat q^{\frac{5}{2}}
+40\hat q^{\frac{7}{2}}
+40\hat q^{\frac{9}{2}}
+{\cal O}(\hat q^{\frac{11}{2}}),
\\
{\cal I}^{{\rm ABJM}(1/2)}_{N=2}|_{\hat u=1}
&=10\hat q
+65\hat q^2
+220\hat q^3
+455\hat q^4
+1060\hat q^5
+1645\hat q^6
+{\cal O}(\hat q^{\frac{13}{2}}),
\\
{\cal I}^{{\rm ABJM}(1/2)}_{N=3}|_{\hat u=1}
&=20\hat q^{\frac{3}{2}}
+164\hat q^{\frac{5}{2}}
+780\hat q^{\frac{7}{2}}
+2500\hat q^{\frac{9}{2}}
+6300\hat q^{\frac{11}{2}}
+15720\hat q^{\frac{13}{2}}
\nonumber\\&
+30496\hat q^{\frac{15}{2}}
+{\cal O}(\hat q^8).
\end{align}

On the gravity side we need to consider wrapped M5-brane with $B=1$.
Because $B$ is $\ZZ_2$-valued $B=+1$ and $B=-1$ are identified,
and all four configurations $z_a=0$ ($a=1,2,3,4$) contribute to the index;
\begin{align}
{\cal I}^{{\rm grav}(1/2)}_N
={\cal I}_{\rm KK}^{\ZZ_2}\sum_{a=1}^4\hat q^{\frac{1}{2}N}\hat u_a^N\Pexp{\cal P}_2 i_{z_a=0}^{\rm M5}.
\label{grav12}
\end{align}
The results for $N=1,2,3$ are
\begin{align}
{\cal I}^{{\rm grav}(1/2)}_{N=1}|_{\hat u=1}
&=4\hat q^{\frac{1}{2}}
+16\hat q^{\frac{3}{2}}
+20\hat q^{\frac{5}{2}}
+40\hat q^{\frac{7}{2}}
-1500\hat q^{\frac{9}{2}}
+{\cal O}(\hat q^5),
\\
{\cal I}^{{\rm grav}(1/2)}_{N=2}|_{\hat u=1}
&=10\hat q
+65\hat q^2
+220\hat q^3
+455\hat q^4
+1060\hat q^5
-7210\hat q^6
+{\cal O}(\hat q^{\frac{13}{2}}),
\\
{\cal I}^{{\rm grav}(1/2)}_{N=3}|_{\hat u=1}
&=20\hat q^{\frac{3}{2}}
+164\hat q^{\frac{5}{2}}
+780\hat q^{\frac{7}{2}}
+2500\hat q^{\frac{9}{2}}
+6300\hat q^{\frac{11}{2}}
+15720\hat q^{\frac{13}{2}}
\nonumber\\&
-12008\hat q^{\frac{15}{2}}
+{\cal O}(\hat q^8).
\end{align}

In all cases the leading term is of order $\hat q^{\frac{1}{2}N}$, and there is no
tachyonic shift.
This is because the $\ZZ_2$ projection removes the tachyonic term from the single-particle index.
This is consistent with the fact that the branes are wrapped on topologically
non-trivial cycles.
The error between the ABJM index and
(\ref{grav12}) appears at $\hat q^{\frac{1}{2}(3N+6)}$.
This is consistent with the fact that only brane configuration
with odd $n$ contribute to the index of the $B=1$ sector
and the error is due to $n=3$ configurations.

\subsubsection{$k=3$}
The $\ZZ_k$ orbifolding with $k\geq3$ breaks the ${\cal N}=8$ supersymmetry
down to ${\cal N}=6$.

We consider $k=3$ case and there are three sectors specified by $B\in\ZZ_3$.
Let us first consider the $B=0$ sector.
The ABJM index is given for $N=1,2,3$ as follows.
\begin{align}
{\cal I}^{{\rm ABJM}(0/3)}_{N=1}|_{\hat u=1}
&=1
+4\hat q
+8\hat q^{\frac{3}{2}}
+\hat q^2
+{\cal O}(\hat q^{\frac{5}{2}}),
\\
{\cal I}^{{\rm ABJM}(0/3)}_{N=2}|_{\hat u=1}
&=1
+4\hat q
+8\hat q^{\frac{3}{2}}
+12\hat q^2
+40\hat q^{\frac{5}{2}}
+58\hat q^3
+{\cal O}(\hat q^{\frac{7}{2}}),
\\
{\cal I}^{{\rm ABJM}(0/3)}_{N=3}|_{\hat u=1}
&=1
+4\hat q
+8\hat q^{\frac{3}{2}}
+12\hat q^2
+40\hat q^{\frac{5}{2}}
+82\hat q^3
+132\hat q^{\frac{7}{2}}
+303\hat q^4
\nonumber\\&
+{\cal O}(\hat q^{\frac{9}{2}}).
\end{align}
Let us compare these with the Kaluza-Klein index
\begin{align}
{\cal I}_{\rm KK}^{\ZZ_3}|_{\hat u=1}
&=1
+4\hat q
+8\hat q^{\frac{3}{2}}
+12\hat q^2
+40\hat q^{\frac{5}{2}}
+82\hat q^3
+132\hat q^{\frac{7}{2}}
+348\hat q^4
+{\cal O}(\hat q^{\frac{9}{2}}).
\end{align}
We find the corrections at $\hat q^{\frac{1}{2}(2N+2)}$.
We can interpret these corrections as the contributions of
brane configurations with $n=2$
consisting of a brane with $B=+1$ and another brane with $B=-1$.

Next, let us consider baryonic sectors with $B=\pm 1$.
These two sectors are related by the charge conjugation symmetry $B\rightarrow -B$
we focus only on the $B=+1$ sector.
The ABJM index is given as follows for $N=1,2,3$.
\begin{align}
{\cal I}^{{\rm ABJM}(1/3)}_{N=1}|_{\hat u=1}
&=2\hat q^{\frac{1}{2}}
+3\hat q
+4\hat q^{\frac{3}{2}}
+9\hat q^2
+{\cal O}(\hat q^{\frac{5}{2}}),
\\
{\cal I}^{{\rm ABJM}(1/3)}_{N=2}|_{\hat u=1}
&=3\hat q
+6\hat q^{\frac{3}{2}}
+14\hat q^2
+32\hat q^{\frac{5}{2}}
+51\hat q^3
+{\cal O}(\hat q^{\frac{7}{2}}),
\\
{\cal I}^{{\rm ABJM}(1/3)}_{N=3}|_{\hat u=1}
&=4\hat q^{\frac{3}{2}}
+9\hat q^2
+24\hat q^{\frac{5}{2}}
+65\hat q^3
+126\hat q^{\frac{7}{2}}
+215\hat q^4
+{\cal O}(\hat q^{\frac{9}{2}}).
\end{align}
On the gravity side we take only two single-wrapping configurations $z_1=0$ and $z_2=0$ into account
because the other two carry $B=-1$.
\begin{align}
{\cal I}^{{\rm grav}(1/3)}_N
={\cal I}_{\rm KK}^{\ZZ_3}\sum_{a=1}^2 \hat q^{\frac{1}{2}N}\hat u_a^N\Pexp{\cal P}_3 i_{z_a=0}^{\rm M5}.
\label{grav13}
\end{align}
The results for $N=1,2,3$ are
\begin{align}
{\cal I}^{{\rm grav}(1/3)}_{N=1}|_{\hat u=1}
&=2\hat q^{\frac{1}{2}}
+3\hat q
+4\hat q^{\frac{3}{2}}
-\hat q^2
+{\cal O}(\hat q^{\frac{5}{2}}),
\\
{\cal I}^{{\rm grav}(1/3)}_{N=2}|_{\hat u=1}
&=3\hat q
+6\hat q^{\frac{3}{2}}
+14\hat q^2
+32\hat q^{\frac{5}{2}}
+36\hat q^3
+{\cal O}(\hat q^{\frac{7}{2}}),
\\
{\cal I}^{{\rm grav}(1/3)}_{N=3}|_{\hat u=1}
&=4\hat q^{\frac{3}{2}}
+9\hat q^2
+24\hat q^{\frac{5}{2}}
+65\hat q^3
+126\hat q^{\frac{7}{2}}
+194\hat q^4
+{\cal O}(\hat q^{\frac{9}{2}}).
\end{align}
We find errors at $\hat q^{\frac{1}{2}(2N+2)}$.
We can interpret them as the contribution of
$n=2$ configurations with $B=-2\approx +1$.

\section{6d ${\cal N}=(2,0)$ superconformal theories}\label{an6d.sec}
In this section we consider
six-dimensional ${\cal N}=(2,0)$ superconformal
theories realized on a stack of M5-branes.
The gravity dual is M-theory in $AdS_7\times \bm{S}^4$.
The $AdS_7$ radius $\check L$, the $\bm{S}^4$ radius $\check r$,
and the M2-brane tension $T_{\rm M2}$ satisfy
the following relations similar to
(\ref{d3brane}) and (\ref{ads4params}):
\begin{align}
\frac{\check L}{\check r}=2,\quad
V_2\check r^3T_{\rm M2}=N.
\label{ads7params}
\end{align}
The ratio $\check L/\check r=2$ gives the dimension of a free scalar field in six-dimension,
and the second relation suggests that wrapped M2-branes in $\bm{S}^4$ are
responsible for the finite-$N$ corrections in
the superconformal index.

\subsection{Superconformal index}
The six-dimensional
${\cal N}=(2,0)$ superconformal algebra is
$\check{\cal A}:=osp(8^*|4)$,
whose bosonic subalgebra is
\begin{align}
so(2,6)\times so(5)\subset \check{\cal A}.
\end{align}
There are six Cartan generators:
\begin{align}
\check H,\quad
\check J_{12},\quad
\check J_{34},\quad
\check J_{56},\quad
\check R_{12},\quad
\check R_{34}.
\end{align}
To define the superconformal index we
need to choose one complex supercharge $\check{\cal Q}$ carrying specific Cartan charges.
We take the one with the quantum numbers
\begin{align}
\check{\cal Q}:(\check H,\check J_{12},\check J_{34},\check J_{56};\check R_{12},\check R_{34})
=(
+\tfrac{1}{2},
-\tfrac{1}{2},
-\tfrac{1}{2},
-\tfrac{1}{2};
+\tfrac{1}{2},
+\tfrac{1}{2}
).
\label{qcheck}
\end{align}
The subalgebra that keeps $\check{\cal Q}$ intact is
\begin{align}
\check{\cal B}\times u(1)_{\check\Delta},
\end{align}
where $\check{\cal B}=osp(6|2)$ is the superalgebra
whose bosonic subalgebra is $su(1,3)\times su(2)\subset\check{\cal B}$.
The central factor $u(1)_{\check\Delta}$
is generated by
\begin{align}
\check\Delta\equiv \{\check{\cal Q},\check{\cal Q}^\dagger\}
=\check H-(\check J_{12}+\check J_{34}+\check J_{56})-2(\check R_{12}+\check R_{34}).
\end{align}
We define the superconformal index associated with the BPS bound $\check\Delta\geq 0$ as the $\check{\cal B}$
character by%
\footnote{The fugacities $\check q$, $\check y_i$, and $\check u$ are related
to those used in Section 3 of \cite{Bhattacharya:2008zy} by
$\check q=x^3$,
$\check y_1=y_1$,
$\check y_2=y_1^{-1}y_2$,
$\check y_3=y_2^{-1}$, and
$\check u=z^{\frac{1}{2}}$}
\begin{align}
{\cal I}(\check q,\check y_a,\check u)
=\tr[(-1)^F\check x^{\check\Delta}
\check q^{\check H+\frac{1}{3}(\check J_{12}+\check J_{34}+\check J_{56})}
\check y_1^{\check J_{12}}
\check y_2^{\check J_{34}}
\check y_3^{\check J_{56}}
\check u^{\check R_{12}-\check R_{34}}
],\quad
\check y_1\check y_2\check y_3=1.
\label{index6d}
\end{align}
Due to the Bose-Fermi degeneracy for $\check\Delta>0$
this does not depend on $\check x$.

\subsection{Wrapped M2-branes}
Let ${\cal I}_N^{(2,0)}$ be the superconformal index
of the theory realized on the stack of $N$ M5-branes.
The large-$N$ limit ${\cal I}_{N=\infty}^{(2,0)}$ is
given by the Kaluza-Klein index of $AdS_7\times \bm{S}^4$.
It is given by
${\cal I}_{\rm KK}=\Pexp i_{\rm KK}$ with
the single-particle index \cite{Bhattacharya:2008zy}
\begin{align}
i_{\rm KK}
&=\frac{
\check q^2\chi_1(\check u)
-\check q^{\frac{8}{3}}\chi_{[0,1]}(\check y)
+\check q^{\frac{16}{3}}\chi_{[1,0]}(\check y)
-\check q^6\chi_1(\check u)
}{(1-\check u\check q^2)(1-\check u^{-1}\check q^2)
(1-\check y_1\check q^{\frac{4}{3}})
(1-\check y_2\check q^{\frac{4}{3}})
(1-\check y_3\check q^{\frac{4}{3}})
},\label{kkspi}
\end{align}
where
$\chi_m(\check u)$ is the $su(2)$ character of the spin $m/2$ representation
\begin{align}
\chi_m(\check u)
=\frac{\check u^{m+1}-\check u^{-m-1}}{\check u-\check u^{-1}}
=\check u^m+\cdots +\check u^{-m},
\end{align}
and
$\chi_{[a,b]}(\check y)$ is the $su(3)$ character of the representation with
Dynkin labels $[a,b]$.
$\chi_{[1,0]}$ for the fundamental representation and $\chi_{[0,1]}$ for the anti-fundamental
representation are
\begin{align}
\chi_{[1,0]}(\check y)=\check y_1+\check y_2+\check y_3,\quad
\chi_{[0,1]}(\check y)=\check y_1^{-1}+\check y_2^{-1}+\check y_3^{-1}.
\end{align}

For the theory on a finite number of M5-branes
we propose the formula
\begin{align}
{\cal I}^{(2,0)}_N={\cal I}_{\rm KK}
\left(1+\sum_C{\cal I}_C^{\rm M2}\right).
\label{m2corrections}
\end{align}
${\cal I}_C^{\rm M2}$ is the contribution
of an M2-brane configuration $C$.
The sum of $C$ runs over
representative configurations,
which are determined shortly
in a parallel way to the three-dimensional case.
Let us introduce Cartesian coordinates
$x_1,\ldots,x_5$ and describe $\bm{S}^4$ by
$\sum_{a=1}^5x_a^2=1$.
We also introduce the complex coordinates
\begin{align}
z_1=x_1+ix_2,\quad
z_2=x_3+ix_4.
\end{align}
The subalgebra $su(2)\subset so(5)$ of the R-symmetry commuting with $\check{\cal Q}$
transforms these complex coordinates as a doublet.
For a rigid M2-brane wrapped on a large $\bm{S}^2$ in $\bm{S}^4$ to preserve the supersymmetry $\check{\cal Q}$,
the M2-brane worldvolume
must be given by the holomorphic equation \cite{Mikhailov:2000ya}
\begin{align}
a_1z_1+a_2z_2=0,
\end{align}
where $(a_1,a_2)$ are homogeneous coordinates
of the moduli space $\PP^1$ of the rigid brane.
Due to the coupling to the background flux the wave function $\Psi$ of the rigid brane is
a section of ${\cal O}(N)$ line bundle over $\PP^1$.
Namely, $\Psi$ can be given as a homogeneous polynomial of $(a_1,a_2)$ of degree $N$.
There are $N+1$ such linearly independent polynomials
belonging to the ($N+1$)-dimensional representation
of $su(2)$ acting on $\PP^1$.
The corresponding index is
\begin{align}
\check q^{2N}\chi_N(\check u)
=\frac{\check q^{2N}\check u^N}{1-\check u^{-2}}
+\frac{\check q^{2N}\check u^{-N}}{1-\check u^2}.
\end{align}
As in the case of wrapped M5-branes the two terms
are interpreted as the contribution of
two representative configurations
of M2-brane, $z_1=0$ and $z_2=0$, respectively.
The general representative configurations are given in the form
\begin{align}
C:z_1^{n_1}z_2^{n_2}=0,\quad
n_1,n_2\in\ZZ_{\geq0},\quad
(n_1,n_2)\neq(0,0),
\label{repr5}
\end{align}
and the corresponding contribution ${\cal I}^{\rm M2}_C$ is given by
\begin{align}
{\cal I}^{\rm M2}_C=\check q^{2nN}\check u^{(n_1-n_2)N}{\cal I}_C^{\rm excitations},
\end{align}
where $n=n_1+n_2$.
For $C$ with $n\geq2$ it is difficult to calculate ${\cal I}_C^{\rm excitations}$,
while for $n=1$ configurations $z_a=0$ ($a=1,2$) the theory on the wrapped brane is free and given by
${\cal I}_{z_a=1}^{\rm excitations}=\Pexp i^{\rm M2}_{z_a=0}$,
where $i^{\rm M2}_{z_a=0}$ is the single-particle index on an M2-brane wrapped on $z_a=0$.

Let us consider
an M2-brane wrapped on $\bm{S}^2\subset\bm{S}^4$ on $z_1=0$.
Among $32$ supercharges only $16$ that commute with
\begin{align}
\check Z=\check H-\check R_{12}
\end{align}
are
preserved by the wrapped brane.
The superconformal algebra $\check{\cal A}$
is broken to
\begin{align}
so(2)_{\check Z}\times\check {\cal C},\quad
\check{\cal C}=su(4|2),
\label{m2unbroken}
\end{align}
where $so(2)_{\check Z}$
is the central factor generated by $\check Z$.
The bosonic subalgebra $su(4)\times su(2)\times u(1)\subset\check{\cal C}$ is
generated by
\begin{align}
\check J_{ij}\quad(i,j=1,\ldots,6),\quad
\check R_{ab}\quad(a,b=3,4,5),\quad
\check C\equiv \check H-2\check R_{12}.
\end{align}
As is explained in the last section
this is isomorphic to the symmetry preserved by
a wrapped M5-brane in (\ref{m5unbroken}).
By using the isomorphism map (\ref{isomorphism}), we can
obtain $i^{\rm M2}_{z_1=0}$ from $i^{\rm M2}_{\rm bdr}$ in (\ref{n1m2index}) by a simple variable change.
The inverse of (\ref{ads7toads4}) is
\begin{align}
\hat q=\check q\check u^{-\frac{1}{2}},\quad
\hat u_1=\check q^{\frac{5}{6}}\check y_1\check u^{\frac{1}{4}},\quad
\hat u_2=\check q^{\frac{5}{6}}\check y_2\check u^{\frac{1}{4}},\quad
\hat u_3=\check q^{\frac{5}{6}}\check y_3\check u^{\frac{1}{4}},\quad
\hat u_4=\check q^{-\frac{5}{2}}\check u^{-\frac{3}{4}},
\label{fugrelation}
\end{align}
and by substituting these relations
into (\ref{n1m2index})
we obtain
\begin{align}
i^{\rm M2}_{z_1=0}
&=\frac{
\check q^{-2}\check u^{-1}
-\check q^{\frac{2}{3}}\check u^{-1}\chi_{[0,1]}(\check y)
+\check q^{\frac{4}{3}}\chi_{[1,0]}(\check y)
-\check q^4
}{1-\check q^2\check u^{-1}}.
\label{m2isp}
\end{align}
The index $i^{\rm M2}_{z_2=0}$ for the other configuration $z_2=0$ is
obtained from (\ref{m2isp}) by the Weyl reflection $\check u\rightarrow \check u^{-1}$.

It is of course possible to calculate the index directly by the
mode expansion of fields on the wrapped brane.
We show the results for scalar fields in Table \ref{m2scalarmodes}.
\begin{table}
  \caption{Scalar modes on an M2-brane wrapped on $z_1=0$. $\ell=0,1,2,\ldots$ is the orbital angular momentum in $\bm{S}^2$. States with $(R_{12},R_{34})=(-1,\ell)$ saturate the BPS bound
  $\check H\geq2(\check R_{12}+\check R_{34})$.}
  \label{m2scalarmodes}
\begin{center}
\begin{tabular}{cccc}
\hline
\hline
$so(6)$ & $\check R_{12}$ & $\check R_{34}$ & $\check H$ \\
\hline
$\bm{6}$ & $0$ & $-\ell\sim\ell$ & $2\ell+1$ \\
$\bm{1}$ & $+1$ & $-\ell\sim\ell$ & $2\ell+4$ \\
$\bm{1}$ & $-1$ & $-\ell\sim\ell$ & $2\ell-2$ \\
\hline
\end{tabular}
\end{center}
\end{table}
There is one BPS
tachyonic mode with $\check H=-2$ and one BPS zero mode.
These correspond to the first two terms in the $\check q$ expansion
of $i^{\rm M2}_{z_1=0}$:
\begin{align}
i^{\rm M2}_{z_1=0}=\frac{1}{\check q^2\check u}+\frac{1}{\check u^2}+\cdots.
\label{firsttwo}
\end{align}

\subsection{Results and consistency check}\label{6dresults.sec}
The formula for the finite-$N$ corrections is
\begin{align}
{\cal I}^{(2,0)}_N={\cal I}_N^{\rm grav}+{\cal O}(\check q^{2(2N+\delta)}),
\end{align}
where ${\cal I}_N^{\rm grav}$ includes the Kaluza-Klein contribution and the
contribution of single wrapping M2-branes,
and the second term is the contribution of multiple wrapping configurations,
which we do not calculate in this paper.
$\delta$ is the tachyonic shift
of configurations with $n=2$.
The explicit form of ${\cal I}_N^{\rm grav}$ is
\begin{align}
{\cal I}^{\rm grav}_N=
{\cal I}_{\rm KK}\left(1
+\check q^{2N}\check u^N\Pexp i^{\rm M2}_{z_1=0}
+\check q^{2N}\check u^{-N}\Pexp i^{\rm M2}_{z_2=0}
\right).
\label{igrav6d}
\end{align}

Let us first consider the $N=1$ case.
In this case, the six-dimensional theory is
the free theory of a single tensor multiplet.
The single-particle index of the tensor multiplet is \cite{Bhattacharya:2008zy}
\begin{align}
i^{\rm M5}_{\rm bdr}=
\frac
{
\check q^2\chi_1(\check u)
-\check q^{\frac{8}{3}}\chi_{[0,1]}(\check y)
+\check q^4
}
{
(1-\check q^{\frac{4}{3}}\check y_1)
(1-\check q^{\frac{4}{3}}\check y_2)
(1-\check q^{\frac{4}{3}}\check y_3)
}.
\label{freetensorindex}
\end{align}
The index for $N=1$ theory is given by ${\cal I}^{(2,0)}_{N=1}=\Pexp i^{\rm M5}_{\rm bdr}$,
and its $\check q$-expansion is
\begin{align}
{\cal I}^{(2,0)}_{N=1}
&=
1+\chi^{\check u}_{1} \check{q}^2-\chi_{[0,1]} \check{q}^{\frac{8}{3}}
+\chi^{\check u}_{1} \chi_{[1,0]} \check{q}^{\frac{10}{3}}
+(\chi^{\check u}_{2}-\chi_{[1,1]}) \check{q}^4
+\chi^{\check u}_{1} (\chi_{[2,0]}-\chi_{[0,1]}) \check{q}^{\frac{14}{3}}
\nonumber\\&
+((\chi^{\check u}_{2}+2) \chi_{[1,0]}-\chi_{[2,1]}) \check{q}^{\frac{16}{3}}
+(\chi^{\check u}_{3}+\chi^{\check u}_{1} (-2 \chi_{[1,1]}+\chi_{[3,0]}-1)) \check{q}^6
\nonumber\\&
+(-(\chi^{\check u}_{2}-2) \chi_{[0,1]}+\chi_{[1,2]}+2 \chi^{\check u}_{2} \chi_{[2,0]}+2 \chi_{[2,0]}-\chi_{[3,1]})
   \check{q}^{\frac{20}{3}}
\nonumber\\&
+(\chi^{\check u}_{3} \chi_{[1,0]}+\chi^{\check u}_{1} (-\chi_{[0,2]}-3 \chi_{[2,1]}+\chi_{[4,0]})) \check{q}^{\frac{22}{3}}
\nonumber\\&
+(\chi^{\check u}_{4}+\chi_{[0,3]}+2 \chi_{[1,1]}+\chi_{[2,2]}-\chi^{\check u}_{2} (\chi_{[1,1]}-2 \chi_{[3,0]}+1)+4 \chi_{[3,0]}-\chi_{[4,1]}-2)
   \check{q}^8
\nonumber\\&
+(\chi^{\check u}_{1} (2 \chi_{[0,1]}-\chi_{[1,2]}+\chi_{[2,0]}-4 \chi_{[3,1]}+\chi_{[5,0]})-\chi^{\check u}_{3} (\chi_{[0,1]}-2 \chi_{[2,0]}))
   \check{q}^{\frac{26}{3}}
\nonumber\\&
+(-2\chi_{[0,2]}+(-\chi^{\check u}_{2}+\chi^{\check u}_{4}-3) \chi_{[1,0]}+\chi_{[1,3]}-3 \chi^{\check u}_{2} \chi_{[2,1]}+2 \chi_{[2,1]}+2 \chi_{[3,2]}
\nonumber\\&\quad
+3 \chi^{\check u}_{2} \chi_{[4,0]}+4 \chi_{[4,0]}-\chi_{[5,1]})
   \check{q}^{\frac{28}{3}}
\nonumber\\&
+(\chi^{\check u}_{5}-\chi^{\check u}_{3} (\chi_{[1,1]}-3 \chi_{[3,0]}+1)+\chi^{\check u}_{1} (\chi_{[0,3]}+6 \chi_{[1,1]}-\chi_{[2,2]}+3 \chi_{[3,0]}-5 \chi_{[4,1]}
\nonumber\\&\quad
+\chi_{[6,0]}-1))
   \check{q}^{10}+{\cal O}(\check{q}^{\frac{55}{3}}).
\end{align}
We use the simplified notations $\chi_m^{\check u}=\chi_m(\check u)$ and $\chi_{[a,b]}=\chi_{[a,b]}(\check y)$.
On the other hand,
the formula (\ref{igrav6d}) with $N=1$ gives
\begin{align}
{\cal I}^{\rm grav}_{N=1}
&=1+\chi _1^{\check u}\check{q}^2
 - \chi _{[0,1]}\check{q}^{\frac{8}{3}}
+ \chi _{[1,0]} \chi _1^{\check u}\check{q}^{\frac{10}{3}}
+ \left(\chi _2^{\check u}-\chi _{[1,1]}\right)\check{q}^4
+ \left(\chi _{[2,0]}-\chi _{[0,1]}\right) \chi _1^{\check u}\check{q}^{\frac{14}{3}}
\nonumber\\
&+\left(\chi _{[1,0]} \left(\chi _2^{\check u}+2\right)-\chi _{[2,1]}\right)\check{q}^{\frac{16}{3}}
+\left(\left(-2 \chi _{[1,1]}+\chi _{[3,0]}-1\right) \chi _1^{\check u}+\chi _3^{\check u}\right)\check{q}^6\nonumber\\
&+\left(2 \chi _{[2,0]} \chi _2^{\check u}-\chi _{[0,1]} \left(\chi _2^{\check u}-2\right)+\chi _{[1,2]}+2 \chi _{[2,0]}-\chi _{[3,1]}\right)\check{q}^{\frac{20}{3}}\nonumber\\
&+\left(-\chi _{[0,2]} \chi _1^{\check u}-3 \chi _{[2,1]}\chi _1^{\check u}+\chi _{[4,0]} \chi _1^{\check u}+\chi _{[1,0]} \chi _3^{\check u}\right)\check{q}^{\frac{22}{3}}\nonumber\\
&+\left(2 \chi _{[3,0]} \chi _2^{\check u}-\chi _{[1,1]} \left(\chi _2^{\check u}-2\right)+\chi _{[0,3]}+\chi _{[2,2]}+4 \chi_{[3,0]}-\chi _{[4,1]}-\chi _2^{\check u}+\chi _4^{\check u}-2\right)\check{q}^8\nonumber\\
&+\left(-\chi _{[1,2]} \chi _1^{\check u}+\chi _{[2,0]} \chi _1^{\check u}-4 \chi _{[3,1]} \chi _1^{\check u}+\chi _{[5,0]} \chi _1^{\check u}+\chi _{[0,1]} \left(2 \chi _1^{\check u}-\chi _3^{\check u}\right)+2 \chi _{[2,0]} \chi _3^{\check u}\right)\check{q}^{\frac{26}{3}}\nonumber\\
&+\left(-3 \chi _{[2,1]} \chi _2^{\check u}+3 \chi _{[4,0]} \chi _2^{\check u}+\chi _{[1,0]} \left(-\chi_2^{\check u}+\chi _4^{\check u}-3\right)\right.\nonumber\\
&\left.\qquad-2 \chi _{[0,2]}+\chi _{[1,3]}+2 \chi _{[2,1]}+2 \chi _{[3,2]}+4 \chi _{[4,0]}-\chi _{[5,1]}\right)\check{q}^{\frac{28}{3}}\nonumber\\
&+\big(\left(\chi _{[0,3]}+6 \chi _{[1,1]}-\chi_{[2,2]}+3 \chi _{[3,0]}-5 \chi _{[4,1]}+\chi _{[6,0]}-1\right) \chi _1^{\check u}
\nonumber\\&\quad\quad
-\left(\chi _{[1,1]}-3 \chi _{[3,0]}+1\right) \chi _3^{\check u}\big)\check{q}^{10} +\mathcal{O}(\check{q}^{\frac{32}{3}}).
\end{align}
We find nice agreement.
The error appears at order
$\check q^{10}$.
This means the tachyonic shift $\delta=3$.
Although we have no interpretation of this value of $\delta$,
let us assume that this is $N$-independent as in the 3d and 4d cases.

The first few terms for $N\geq 2$ are
\begin{align}
{\cal I}^{\rm grav}_{N\geq 2}
&=1+\chi _1^{\check u}\check{q}^2
- \chi _{[0,1]}\check{q}^{\frac{8}{3}}
+ \chi _{[1,0]} \chi _1^{\check u}\check{q}^{\frac{10}{3}}
+ \left(2 \chi _2^{\check u}-\chi _{[1,1]}\right)\check{q}^4
+ \left(\chi _{[2,0]}-2 \chi _{[0,1]}\right) \chi _1^{\check u}\check{q}^{\frac{14}{3}}
\nonumber\\
&+ \left(\chi _{[1,0]} \left(2 \chi _2^{\check u}+3\right)-\chi _{[2,1]}\right)\check{q}^{\frac{16}{3}}
+{\cal O}(\check{q}^6).
\label{ngeq2}
\end{align}
The leading finite-$N$ correction given by (\ref{igrav6d}) is $-\chi^{\check{u}}_{N+1}\check{q}^{2(N+1)}$,\footnote{In
\cite{Kim:2013nva}, the finite-$N$ index of the 6d $(2,0)$ theory was studied
from 5d SYM on $\CC\PP^2\times \bm{S}^1$.
Especially, the authors found that the leading correction for $N=2$ is $-q^3 y^3$,
where the fugacities $q$ and $y$ are related to ours by
$\check{q}= q^{\frac{3}{4}}$ and $\check{u}= q^{-\frac{1}{2}} y$.
See (3.65) in \cite{Kim:2013nva}.
This is consistent with our result: $-\chi^{\check{u}}_{3}\check{q}^{6}=-q^3 y^3+\cdots$.}
and the terms in the range shown in (\ref{ngeq2}) is the same as the supergravity approximation ${\cal I}_{\rm KK}$.

The second term $\chi_1^{\check u}\check q^2$ is the contribution of the primary operators
in the free tensor multiplet.
The term $\chi_2^{\check u}\check q^4$ is the contribution of the
stress-tensor multiplet.
The coefficient $2$ of the term suggests that the theory has two stress-energy tensors.
Namely, the system consists of two decoupled theories.
One is the free theory of the tensor multiplet,
and the other is the interacting theory called the $A_{N-1}$ theory.

By removing the contribution of the free tensor multiplet
we obtain the index of the $A_{N-1}$ theory:
\begin{align}
{\cal I}_{A_{N-1}}=\frac{{\cal I}^{(2,0)}_N}{{\cal I}^{(2,0)}_{N=1}}.
\end{align}

Explicit forms of ${\cal I}_{A_{N-1}}$ for small $N$ obtained
by using (\ref{igrav6d}) are as follows.
\begin{align}
{\cal I}_{A_1}&=
1+\chi ^{\check u}_2\check{q}^4 - \chi_{[0,1]} \chi ^{\check u}_1\check{q}^{\frac{14}{3}}+ \chi_{[1,0]} \left(\chi ^{\check u}_2+1\right)\check{q}^{\frac{16}{3}}
- (\chi_{[1,1]}+1) \chi ^{\check u}_1\check{q}^6\nonumber\\
&+\left(\chi_{[2,0]} \left(\chi^{\check u}_2+1\right)+\chi_{[0,1]}\right)\check{q}^{\frac{20}{3}}
-(\chi_{[1,0]}+\chi_{[2,1]}) \chi ^{\check u}_1\check{q}^{\frac{22}{3}} \nonumber\\
&+\left(\chi_{[3,0]} \left(\chi ^{\check u}_2+1\right)+\chi_{[1,1]}+\chi^{\check u}_4\right)\check{q}^8 + \left(-\chi_{[2,0]} \chi ^{\check u}_1-\chi_{[3,1]} \chi ^{\check u}_1-\chi_{[0,1]} \chi ^{\check u}_3\right)\check{q}^{\frac{26}{3}}\nonumber\\
&+\left(\chi_{[4,0]} \left(\chi ^{\check u}_2+1\right)+\chi _{[1,0]} \left(2 \chi ^{\check u}_2+\chi ^{\check u}_4\right)+\chi_{[2,1]}\right)\check{q}^{\frac{28}{3}} \nonumber\\
&+\left(
-\chi_{[1,1]}( \chi ^{\check u}_1+2\chi^{\check u}_3)
-\chi_{[3,0]} \chi ^{\check u}_1
-\chi_{[4,1]} \chi ^{\check u}_1
-2 \chi ^{\check u}_1
-\chi^{\check u}_3
\right)\check{q}^{10}
\nonumber\\&
   +{\cal O}(\check{q}^{\frac{32}{3}}).
\end{align}
\begin{align}
{\cal I}_{A_2}&=
1+\chi ^{\check u}_2\check{q}^4-\chi_{[0,1]} \chi ^{\check u}_1\check{q}^{\frac{14}{3}}+\chi_{[1,0]} \left(\chi ^{\check u}_2+1\right)\check{q}^{\frac{16}{3}}
+\left(\chi ^{\check u}_3-(\chi_{[1,1]}+1) \chi ^{\check u}_1\right)\check{q}^6\nonumber\\
&+\left(\chi_{[2,0]} \left(\chi ^{\check u}_2+1\right)-\chi_{[0,1]} \left(\chi ^{\check u}_2-1\right)\right)\check{q}^\frac{20}{3}\nonumber\\
 &+\left(\chi_{[1,0]} \chi ^{\check u}_3-\chi_{[2,1]} \chi ^{\check u}_1\right)\check{q}^{\frac{22}{3}} +\left(-\chi_{[1,1]} \left(\chi ^{\check u}_2-1\right)+\chi_{[3,0]} \left(\chi ^{\check u}_2+1\right)-\chi ^{\check u}_2+\chi ^{\check u}_4\right)\check{q}^8\nonumber\\
&+\left((\chi_{[2,0]}-\chi_{[0,1]}) \chi^{\check u}_3-\chi_{[3,1]} \chi ^{\check u}_1\right)\check{q}^{\frac{26}{3}}\nonumber\\
&+\left(-\chi_{[2,1]} \left(\chi ^{\check u}_2-1\right)+2 \chi_{[1,0]} \chi ^{\check u}_2+\chi_{[4,0]} \chi ^{\check u}_2+\chi_{[1,0]} \chi ^{\check u}_4+\chi_{[0,2]}+\chi_{[4,0]}\right)\check{q}^{\frac{28}{3}}\nonumber\\
&+\left(
-2 \chi_{[1,1]}(\chi ^{\check u}_1+\chi ^{\check u}_3)
-\chi_{[4,1]}\chi ^{\check u}_1
+\chi_{[3,0]}\chi ^{\check u}_3
-3\chi ^{\check u}_1
-\chi ^{\check u}_3
+\chi ^{\check u}_5
\right)\check{q}^{10}
\nonumber\\
&
  +{\cal O}(\check{q}^{\frac{32}{3}}).
\end{align}
\begin{align}
{\cal I}_{A_3}
&=
1+\chi ^{\check u}_2\check{q}^4-\chi_{[0,1]} \chi ^{\check u}_1\check{q}^{\frac{14}{3}}+\chi_{[1,0]} \left(\chi ^{\check u}_2+1\right)\check{q}^{\frac{16}{3}}
+\left(\chi ^{\check u}_3-(\chi_{[1,1]}+1) \chi ^{\check u}_1\right)\check{q}^6\nonumber\\
&+\left(\chi_{[2,0]} \left(\chi ^{\check u}_2+1\right)-\chi_{[0,1]} \left(\chi ^{\check u}_2-1\right)\right)\check{q}^\frac{20}{3}\nonumber\\
&+\left(\chi_{[1,0]} \chi ^{\check u}_3-\chi_{[2,1]} \chi ^{\check u}_1\right)\check{q}^{\frac{22}{3}} +\left(-\chi_{[1,1]} \left(\chi ^{\check u}_2-1\right)+\chi_{[3,0]} \left(\chi ^{\check u}_2+1\right)-\chi ^{\check u}_2+2 \chi ^{\check u}_4\right)\check{q}^8\nonumber\\
&+\left((\chi_{[2,0]}-2 \chi_{[0,1]}) \chi^{\check u}_3-\chi_{[3,1]} \chi ^{\check u}_1\right)\check{q}^{\frac{26}{3}} \nonumber\\
&+\left(-\chi_{[2,1]} \left(\chi ^{\check u}_2-1\right)+3 \chi_{[1,0]} \chi ^{\check u}_2+\chi_{[4,0]} \chi ^{\check u}_2+2 \chi_{[1,0]} \chi^{\check u}_4+\chi_{[0,2]}+\chi_{[4,0]}\right)\check{q}^{\frac{28}{3}} \nonumber\\
&+\left(
- \chi_{[1,1]}(2\chi ^{\check u}_1+3\chi ^{\check u}_3)
-\chi_{[4,1]}\chi ^{\check u}_1
+\chi_{[3,0]}\chi ^{\check u}_3
-3\chi ^{\check u}_1
-2\chi ^{\check u}_3
+\chi ^{\check u}_5
\right)\check{q}^{10}
\nonumber\\
&
+\mathcal{O}(\check{q}^{\frac{32}{3}}).
\end{align}
\begin{align}
{\cal I}_{A_{\geq 4}}
&=
1+\chi ^{\check u}_2\check{q}^4-\chi_{[0,1]} \chi ^{\check u}_1\check{q}^{\frac{14}{3}}+\chi_{[1,0]} \left(\chi ^{\check u}_2+1\right)\check{q}^{\frac{16}{3}}
+\left(\chi ^{\check u}_3-(\chi_{[1,1]}+1) \chi ^{\check u}_1\right)\check{q}^6\nonumber\\
&+\left(\chi_{[2,0]} \left(\chi ^{\check u}_2+1\right)-\chi_{[0,1]} \left(\chi ^{\check u}_2-1\right)\right)\check{q}^\frac{20}{3}\nonumber\\
&+\left(\chi_{[1,0]} \chi ^{\check u}_3-\chi_{[2,1]} \chi ^{\check u}_1\right)\check{q}^{\frac{22}{3}} +\left(-\chi_{[1,1]} \left(\chi ^{\check u}_2-1\right)+\chi_{[3,0]} \left(\chi ^{\check u}_2+1\right)-\chi ^{\check u}_2+2 \chi ^{\check u}_4\right)\check{q}^8\nonumber\\
&+\left((\chi_{[2,0]}-2 \chi_{[0,1]}) \chi^{\check u}_3-\chi_{[3,1]} \chi ^{\check u}_1\right)\check{q}^{\frac{26}{3}} \nonumber\\
&+\left(-\chi_{[2,1]} \left(\chi ^{\check u}_2-1\right)+3 \chi_{[1,0]} \chi ^{\check u}_2+\chi_{[4,0]} \chi ^{\check u}_2+2 \chi_{[1,0]} \chi^{\check u}_4+\chi_{[0,2]}+\chi_{[4,0]}\right)\check{q}^{\frac{28}{3}} \nonumber\\
&+\left(
- \chi_{[1,1]}(2\chi ^{\check u}_1+3\chi ^{\check u}_3)
-\chi_{[4,1]}\chi ^{\check u}_1
+\chi_{[3,0]}\chi ^{\check u}_3
-3\chi ^{\check u}_1
-2\chi ^{\check u}_3
+2\chi ^{\check u}_5
\right)\check{q}^{10}
\nonumber\\
&
+\mathcal{O}(\check{q}^{\frac{32}{3}}).
\end{align}
We gave the above $\check q$-expansion
up to $\check q^{10}$ terms.
The error in ${\cal I}_{A_{N-1}}$ estimated with $\delta=3$ is $\check q^{2(2N+3)}$,
and all terms shown above are expected to be correct.

As far as we are aware there are no explicit results
in the literature which can be compared with these results.
As a consistency check, let us expand these results by
indices of superconformal representations.
It is guaranteed by construction that (\ref{igrav6d}) can be expanded
by characters of the bosonic subalgebra $su(3)\times su(2)$.
However, it is non-trivial if it can be expanded by the
indices of superconformal representations.
The results are as follows.
\begin{align}
{\cal I}_{A_1}&
=1
+{\cal D}[2,0]
+{\cal D}[4,0]
+{\cal B}[2,0]_0
+{\cal O}(\check q^{\frac{32}{3}}),
\\
{\cal I}_{A_2}
&=1
+{\cal D}[2,0]
+{\cal D}[3,0]
+{\cal D}[4,0]
+{\cal D}[0,4]
+{\cal B}[2,0]_0
+{\cal D}[5,0]
\nonumber\\&
+{\cal D}[3,2]
+{\cal O}(\check q^{\frac{32}{3}}),
\\
{\cal I}_{A_3}
&=1
+{\cal D}[2,0]
+{\cal D}[3,0]
+2{\cal D}[4,0]
+{\cal D}[0,4]
+{\cal B}[2,0]_0
+{\cal D}[5,0]
\nonumber\\&
+{\cal D}[3,2]
+{\cal D}[1,4]
+{\cal O}(\check q^{\frac{32}{3}}),
\\
{\cal I}_{A_{\geq4}}
&=1
+{\cal D}[2,0]
+{\cal D}[3,0]
+2{\cal D}[4,0]
+{\cal D}[0,4]
+{\cal B}[2,0]_0
+2{\cal D}[5,0]
\nonumber\\&
+{\cal D}[3,2]
+{\cal D}[1,4]
+{\cal O}(\check q^{\frac{32}{3}}).
\end{align}
See Appendix \ref{irrrep}
for the index of each irreducible representation.
We exploited the notation for representations used in \cite{Beem:2015aoa}
to denote the corresponding indices.
These results support the correctness of the formula (\ref{igrav6d}).
In addition, the expansion of ${\cal I}_{A_1}$ seems to be exceptionally simple.
In particular, as was pointed out in \cite{Beem:2015aoa}
the ${\cal D}[0,4]$ representation is absent in the $A_1$ theory.

\subsection{Schur-like index}\label{schur_like_limit.sec}
As shown in (\ref{repr5}) a generic representative configuration
consists of M2-branes wrapped on two cycles $z_1=0$ and $z_2=0$.
We can simplify the problem by taking a special limit in which only one of these two cycles,
say, $z_1=0$,
contributes to the index.
For M2-branes wrapped on $z_2=0$ not to contribute to the index
we need to tune the fugacities so that an extra supersymmetry
which is broken by the
M2-brane wrapped on $z_2=0$
is preserved by the
definition of the index (\ref{index6d}).

The single particle index $i_{z_2=0}^{\rm M2}$ includes
$-\check q^{\frac{2}{3}}\check u\chi_{[0,1]}(\check y)$,
which is the Weyl reflection of the second term in the numerator of (\ref{m2isp}),
and it consists of three terms
\begin{align}
-\check q^{\frac{2}{3}}\check u\chi_{[0,1]}(\check y)
=
-\check q^{\frac{2}{3}}\check u\check y_1^{-1}
-\check q^{\frac{2}{3}}\check u\check y_2^{-1}
-\check q^{\frac{2}{3}}\check u\check y_3^{-1}.
\label{threeterms}
\end{align}
These three terms correspond to Nambu-Goldstone fermions associated with
the breaking of supersymmetry due to the presence of the wrapped brane.
Let us focus on the first term corresponding to
the supercharge $\check{\cal Q}'$ with the quantum numbers
\footnote{The $\ZZ_k$ symmetry (\ref{zkorbi}) acts on the first two terms and the last term in different ways
and this causes inequality between the third one and the others.
We should not take the third term to define the Schur-like limit
because the corresponding supercharge is non-perturbative in the sense that
it is not manifest in the ABJM Lagrangian and is generated dynamically.}
\begin{align}
\check{\cal Q}':(\check H,\check J_{12},\check J_{34},\check J_{56},\check R_{12},\check R_{34})
=(
+\tfrac{1}{2},
-\tfrac{1}{2},
+\tfrac{1}{2},
+\tfrac{1}{2},
+\tfrac{1}{2},
-\tfrac{1}{2}).
\end{align}
To make the definition of the index (\ref{index6d})
respect this supercharge we impose the following condition on the fugacities.
\begin{align}
\check q^{\frac{2}{3}}\check u\check y_1^{-1}=1.
\label{schurlike}
\end{align}
Then the first term in (\ref{threeterms}) becomes $-1$, and
its plethystic exponential vanishes.
As the result, only configurations consisting of M2-branes wrapped on $z_1=0$
contribute to the index.
We adopt the following parametrization of fugacities
satisfying (\ref{schurlike}) (and $\check y_1\check y_2\check y_3=1$).
\begin{align}
\check q  =\check q'\check x',\quad
\check y_1=\check q'^{\frac{2}{3}}\check x'^{-\frac{4}{3}},\quad
\check y_2=\check q'^{-\frac{1}{3}}\check x'^{\frac{2}{3}}\check y,\quad
\check y_3=\check q'^{-\frac{1}{3}}\check x'^{\frac{2}{3}}\check y^{-1},\quad
\check u=\check x'^{-2}.
\label{schurlimit}
\end{align}
New fugacities $\check q'$, $\check x'$, $\check y$ are unconstrained variables.
With this specialization the index (\ref{index6d}) becomes
\begin{align}
\wt{\cal I}(\check q',\check y)
=\tr[(-1)^F\check x^{\check\Delta}
\check x'^{\check\Delta'}
\check q'^{\check H+\check J_{12}}
\check y^{\check J_{34}-\check J_{56}}
],
\label{schur6d}
\end{align}
where
\begin{align}
\check\Delta'=\{\check{\cal Q}',\check{\cal Q}'^\dagger\}
=\check H-(\check J_{12}-\check J_{34}-\check J_{56})-2(\check R_{12}-\check R_{34}).
\end{align}
(\ref{schur6d}) is nothing but the Schur-like index studied in \cite{Beem:2014kka}.
\footnote{The fugacities in this paper are related to those in
\cite{Beem:2014kka} by $\check q'=q^{\frac{1}{2}}$ and $\check y=s$.}
In fact, the analytic result of the index for M5-brane theories
was obtained from
five-dimensional $U(N)$ SYM \cite{Kim:2013nva,Beem:2014kka}:
\begin{align}
\wt{\cal I}_N^{(2,0)}
&
=\Pexp\left[\frac{\check q'^2+\check q'^4+\cdots +\check q'^{2N}}{1-\check q'^2}\right]
=\prod_{k=1}^N\prod_{m=0}^\infty\frac{1}{1-\check q'^{2(k+m)}}
\nonumber\\&
=\wt{\cal I}_{N=\infty}^{(2,0)}
\prod_{k=0}^\infty\prod_{m=0}^\infty(1-\check q'^{2N}\check q'^{2(k+m+1)}).
\label{6danalytic}
\end{align}
By expanding this with respect to $\check q'^{2N}$ we obtain
\begin{align}
\wt{\cal I}_N^{(2,0)}
&=
\wt{\cal I}_{N=\infty}^{(2,0)}
\left(1+\sum_{n=1}^\infty \check q'^{2nN}F_n(\check q')\right),
\label{6danalytic2}
\end{align}
where $F_n(\check q')$ are rational functions of $\check q'$.
The functions for $n=1,2,3$ are
\begin{align}
F_1(\check q')&=\frac{-\check q'^2}{(1-\check q'^2)^2}
=-\check q'^2-2\check q'^4-3\check q'^6-\cdots,\\
F_2(\check q')&=\frac{2\check q'^6}{(1-\check q'^2)^2(1-\check q'^4)^2}
=2\check q'^6+4\check q'^8+10\check q'^{10}+\cdots,\label{f2}\\
F_3(\check q')&=\frac{-\check q'^{10}-4\check q'^{12}-\check q'^{14}}{(1-\check q'^2)^2(1-\check q'^4)^2(1-\check q'^6)^2}
=-\check q'^{10}-6\check q'^{12}-14\check q'^{14}-\cdots.\label{f3}
\end{align}

Let us compare
(\ref{6danalytic2}) with
the hypothetical relation (\ref{m2corrections}), which reduces in the Schur-like limit to the following relation:
\begin{align}
\wt{\cal I}_N^{(2,0)}(\check q',\check y)
=\wt{\cal I}_{\rm KK}\left(1+\sum_{n=1}^\infty\check q'^{2nN}\wt{\cal I}^{\rm M2}_n(\check q',\check y)\right),
\end{align}
where
$\wt{\cal I}^{\rm M2}_n$ is the Schur-like index of the theory
realized on a stack of $n$ M2-branes wrapped around the cycle $z_1=0$.
The agreement in the large-$N$ limit is easily confirmed:
\begin{align}
\wt{\cal I}_{N=\infty}^{(2,0)}=\Pexp\frac{\check q'^2}{(1-\check q'^2)^2}
=\wt{\cal I}_{\rm KK}.
\end{align}
The agreement of finite-$N$ corrections requires
\begin{align}
\wt{\cal I}^{\rm M2}_n(\check q',\check y)
=F_n(\check q')\quad
n=1,2,3,\ldots.
\label{im2n}
\end{align}
For $n=1$, the single-wrapping contribution,
we can easily confirm (\ref{im2n}) by using the Schur-like limit of
$i^{\rm M2}_{z_1=0}$ in (\ref{m2isp})
\begin{align}
\wt i^{\rm M2}_{z_1=0}=\frac{1}{\check q'^2}+\check q'^2.
\end{align}
For $n\geq2$ we expect that $F_n$ is the index of the ABJM theory realized on ${\bm S}^2\subset{\bm S}^4$.
It is straightforward to write down the integral form if
the index.
A non-trivial point is how we should choose the integration contours.
Although at present we have not completely understood it
we found that with a certain prescription
we can reproduce the first few terms in $F_2$ and $F_3$.
See Appendix \ref{abjmformula.sec} for details.

\section{Summary and Discussions}\label{disc.sec}
In this paper
we investigated the superconformal index of theories on M2-branes and M5-branes
using the AdS/CFT correspondence.
We proposed formulas that give finite-$N$ corrections to the superconformal indices
of these theories as the contribution of wrapped M-branes.
We only included single-wrapping brane configurations,
and the contributions of multiple branes are left for future work.

For M2-brane theories we proposed the formula (\ref{igravm2}).
We compared the results with the
results of direct calculation using the ABJM theory.
The results of the comparison are summarized in Table \ref{abjmresults}.
\begin{table}[htb]
\caption{Results of the comparison for the ABJM theory.
$k$, $B$, and $N$ are the Chern-Simons level, the baryonic charge, and the rank of the gauge group.
The column ``single'' shows the orders of the correction due to single-wrapping configurations.
The column ``multiple'' shows the orders of the errors due to multiple-wrapping configurations.}\label{abjmresults}
\centering
\begin{tabular}{cccccc}
\hline
\hline
$k$ & $B$ & $N$ & single & multiple \\
\hline
$1$ & - & $1,2,3$ & $\hat q^{\frac{1}{2}(N+1)}$ & $\hat q^{\frac{1}{2}(2N+6)}$ \\
$2$ & $0$ & $1,2,3$ & - & $\hat q^{\frac{1}{2}(2N+2)}$ \\
$2$ & $1$ & $1,2,3$ & $\hat q^{\frac{1}{2}N}$ & $\hat q^{\frac{1}{2}(3N+6)}$ \\
$3$ & $0$ & $1,2,3$ & - & $\hat q^{\frac{1}{2}(2N+2)}$ \\
$3$ & $1$ & $1,2,3$ & $\hat q^{\frac{1}{2}N}$ & $\hat q^{\frac{1}{2}(2N+2)}$ \\
\hline
\end{tabular}
\end{table}
We found complete agreement up to errors due to multiple branes.
It would be difficult to calculate the contribution of multiple wrapping on the
gravity side because we need to deal with multiple M5-branes.
Conversely, it may be possible to obtain some information about
multiple M5-branes from the higher order corrections
in the ABJM index.
For example, we found that $\delta$ does not depend on $N$
at least for $N=1,2,3$.
This may suggest that the theory on multiple M5-brane does not couple to the
background flux.

For M5-brane theories we proposed the formula (\ref{igrav6d}).
We determined the tachyonic shift $\delta=3$ by using the result for $N=1$.
Under the assumption of $N$-independence of $\delta$, the multiple brane contributions
are of order $\check q^{2(2N+3)}$,
and our formula should give correct index below the order.
We showed the explicit form of the index up to order $\check q^{10}$.
As a consistency check we confirmed that the indices can be
decomposed into the contributions of superconformal irreducible
representations.
In particular, the decomposition of $A_1$ theory is exceptionally simple,
and it seems to match the expectation that the $A_1$ theory is the minimal ${\cal N}=(2,0)$
theory.

There are some proposals about the superconformal index of the $(2,0)$ theories.
The index was related to the partition function or index of five-dimensional
supersymmetric Yang-Mills theories in \cite{Kim:2012qf,Kim:2013nva},
and a relation to topological strings was investigated in \cite{Lockhart:2012vp}.
It is important task to compare their results and ours.
As a first step of this task, we confirmed that the first term of the wrapped M2-brane index for $N=2$
is consistent with the result calculated in \cite{Kim:2013nva}.

We also discussed the Schur-like limit of the 6d superconformal index,
for which an analytic formula is known.
We reported the preliminary result that with a certain prescription for pole selection
we could reproduce the first few terms of multiple-wrapping contributions $F_2$ and $F_3$
in (\ref{f2}) and (\ref{f3}).

We constructed the formulas by extending the Weyl's character formula.
It would be interesting to search for an algebraic structure behind our formulas.
If there exist a large algebra which includes creation and annihilation operators of
not only
fluctuation modes on wrapped branes but also
wrapped branes themselves then it might be possible to regard the whole spectrum of
a boundary theory as an irreducible representation of such an algebra.

There are many ways of extension.
We can consider more general 3d and 6d theories whose
gravity duals are known.
For example, it is easy to extend the formula
to Chern-Simons quiver gauge theories realized on M2-branes in toric Calabi-Yau fourfolds
and 6d ${\cal N}=(1,0)$ theories realized on M5-branes in orbifolds.
It is also important problem to derive the explicit formula for the
multiple brane contributions, together with the pole selection rules in the
gauge fugacity integral.
We hope we could return these issues in near future.

\section*{Acknowledgments}
The work of R. A. was supported by the Sasakawa Scientific Research Grant from The Japan Science Society.
\appendix

\section{Index of the ABJM theory}\label{abjmformula.sec}
In Section \ref{abjm.sec} we calculated the superconformal index of the ABJM theory
as the boundary theory.
The index is given by summing up contributions of
monopole sectors, which are labeled by monopole charges
quantized by
\begin{align}
m_\alpha,\wt m_\alpha\in\ZZ+\frac{B}{k}.
\end{align}
The contribution from each monopole sector is given by
\begin{align}
{\cal I}_{m_\alpha,\wt m_\alpha}=&
\frac{1}{(N!)^2}
\prod_{\alpha=1}^N\int\frac{d\zeta_\alpha}{2\pi i\zeta_\alpha}
\prod_{\alpha=1}^N\int\frac{d\wt \zeta_\alpha}{2\pi i\wt \zeta_\alpha}
\nonumber\\
&\times
\frac{\prod_{\alpha,\beta}\hat q^{|m_\alpha-\wt m_\beta|-\frac{1}{2}|m_\alpha-m_\beta|-\frac{1}{2}|\wt m_\alpha-\wt m_\beta|}}
        {\prod_{\alpha=1}^N \zeta_\alpha^{km_\alpha}\wt \zeta_\alpha^{-k\wt m_\alpha}}
\Pexp i,
\label{abjlocalization}
\end{align}
where $i$ is the single-particle index
\begin{align}
i(\hat q,\hat u_i;\zeta_a,\wt \zeta_b)
&=-\sum_{\alpha\neq\beta}\hat q^{|m_\alpha-m_\beta|}\frac{\zeta_\alpha}{\zeta_\beta}
-\sum_{\alpha\neq\beta}\hat q^{|\wt m_\alpha-\wt m_\beta|}\frac{\wt \zeta_\alpha}{\wt \zeta_\beta}
\nonumber\\
&+\sum_{\alpha,\beta=1}^N\frac{\hat q^{|m_\alpha-\wt m_\beta|}}{1-\hat q^2}
\left[\hat q^{\frac{1}{2}}(\hat u_1+\hat u_2)-\hat q^{\frac{3}{2}}(\hat u_3^{-1}+\hat u_4^{-1})\right]\frac{\zeta_\alpha}{\wt \zeta_\beta}
\nonumber\\
&+\sum_{\alpha,\beta=1}^N\frac{\hat q^{|m_\alpha-\wt m_\beta|}}{1-\hat q^2}
\left[\hat q^{\frac{1}{2}}(\hat u_3+\hat u_4)
-\hat q^{\frac{3}{2}}(\hat u_1^{-1}+\hat u_2^{-1})\right]\frac{\wt \zeta_\beta}{\zeta_\alpha}.
\label{abjmspi}
\end{align}
The gauge fugacity integral gives non-vanishing value only if the monopole charges satisfy
\begin{align}
m_{\rm tot}:=\sum_{\alpha=1}^Nm_\alpha=\sum_{\alpha=1}^N\wt m_\alpha.
\end{align}

In Section \ref{an6d.sec} we discussed Schur-like index of the theory on M2-branes wrapped on the
cycle $z_1=0$.
The integral form giving the Schur-like index of each monopole sector $\wt{\cal I}_{m_\alpha,\wt m_\alpha}$
is obtained from (\ref{abjlocalization}) by setting $k=1$ and the variable change
\begin{align}
\hat q=\check q'\check x'^2,\quad
\hat u_1=\check q'^{\frac{3}{2}}\check x'^{-1},\quad
\hat u_2=\check q'^{\frac{1}{2}}\check x'\check y,\quad
\hat u_3=\check q'^{\frac{1}{2}}\check x'\check y^{-1},\quad
\hat u_4=\check q'^{-\frac{5}{2}}\check x'^{-1}.
\end{align}
These are compositions of (\ref{fugrelation}) and (\ref{schurlimit}).
The single-particle index
(\ref{abjmspi})
reduces to
\begin{align}
\wt i(\check q',\check x';\zeta_\alpha,\wt\zeta_\alpha)
=&-\sum_{\alpha\neq\beta}\hat q^{|m_\alpha-m_\beta|}\frac{\zeta_\alpha}{\zeta_\beta}
-\sum_{\alpha\neq\beta}\hat q^{|\wt m_\alpha-\wt m_\beta|}\frac{\wt\zeta_\alpha}{\wt\zeta_\beta}
\nonumber\\
&+\sum_{\alpha,\beta=1}^N\hat q^{|m_\alpha-\wt m_\beta|}\left(\check q'^2\frac{\zeta_\alpha}{\wt\zeta_\beta}
 +\check q'^{-2}\frac{\wt\zeta_\beta}{\zeta_\alpha}\right),
\end{align}
where we leave $\hat q$ to keep the expression simple.
As expected this is $\check y$-independent.
Although the Schur-like index must be $\check x'$-independent
the single-particle index depends on $\check x'$ through $\hat q=\check q'\check x'^2$.
This is because the above formula is derived by deforming the Lagrangian by $\check{\cal Q}$-exact terms,
which does not respect the extra supercharge $\check{\cal Q}'$ used in the definition of the Schur-like index.

If we regard $\wt{\cal I}_{m_a,\wt m_a}$ as a function of $\check q'$ and $\hat q$, we can easily factor out the
$\check q'$-dependence by the replacement
\begin{align}
\zeta_\alpha\rightarrow \check q'^{-1}\zeta_\alpha,\quad
\wt\zeta_\alpha\rightarrow \check q'\wt\zeta_\alpha,
\end{align}
and obtain
\begin{align}
\wt{\cal I}_{m_\alpha,\wt m_\alpha}=\check q'^{2m_{\rm tot}}\times(\mbox{function of $\hat q$}).
\end{align}
Furthermore, the $\check x'$-independence of the Schur-like index
guarantees that the function of $\hat q$ is in fact a $\hat q$-independent constant.

In order to carry out the gauge fugacity integrals we need to choose integration contours.
Although we have not yet completely understood how we should do it,
we found a prescription that reproduces the known results
after some trial and error.
We express the integrand as the expansion
\begin{align}
\sum_{k=0}^\infty\check q'^k \sum_{l}\check x'^l f_{k,l}(\zeta_\alpha,\wt\zeta_\alpha).
\end{align}
Namely, we first expand the integrand with respect to $\check q'$,
and then expand the result with respect to $\check x'$.
The coefficients $f_{k,l}$ are Laurant polynomials of the gauge fugacities.
The integration over gauge fugacities is equivalent to picking up the terms independent of gauge fugacities
form each $f_{k,l}$.
For each monopole sector
the gauge integral leaves only terms of the form $\check q'^{2m_{\rm tot}}\check x'^0$.
We confirmed that by the summation over monopole sectors the first few terms
of $F_2$ and $F_3$ shown in (\ref{f2}) and (\ref{f3}) are reproduced.

\section{Full expressions of the indices in Section \ref{abjm.sec}}\label{fullexp.sec}
\subsection{$k=1$}
The ABJM theory with the Chern-Simons level $k=1$ has ${\cal N}=8$ supersymmetry
and the index is expanded by the $su(4)$ characters.
In the following
we use $su(4)$ characters $\chi_{[a,b,c]}=\chi_{[a,b,c]}(\hat u_a)$ to write down
the $\hat q$-expansion of the index.
The characters of the fundamental representation $[1,0,0]$ and the anti-fundamental representation $[0,0,1]$
are shown in (\ref{fafcharacters}).
\begin{align}
{\cal I}_{\rm KK}
&=1
+ \chi_{[1,0,0]}\hat q^{\frac{1}{2}}
+2\chi_{[2,0,0]}\hat q
+(3\chi_{[3,0,0]}+\chi_{[1,1,0]}-\chi_{[0,0,1]})\hat q^{\frac{3}{2}}
\nonumber\\&
+(5\chi_{[4,0,0]}+2\chi_{[2,1,0]}+2\chi_{[0,2,0]}-2\chi_{[1,0,1]}-1)\hat q^2
+{\cal O}(\hat q^{\frac{5}{2}}).
\end{align}
\begin{align}
{\cal I}^{\rm ABJM}_{N=1}
&=1
+\chi_{[1, 0, 0]}\hat q^{\frac{1}{2}}
+\chi_{[2, 0, 0]}\hat q
+(-\chi_{[0, 0, 1]} + \chi_{[3, 0, 0]})\hat q^{\frac{3}{2}}
\nonumber\\&
+(-1 - \chi_{[1, 0, 1]} + \chi_{[4, 0, 0]})\hat q^2
+(-\chi_{[2, 0, 1]} + \chi_{[5, 0, 0]})\hat q^{\frac{5}{2}}
\nonumber\\&
+( 2\chi_{[0, 1, 0]} - \chi_{[3, 0, 1]} + \chi_{[6, 0, 0]})\hat q^3
+( 2\chi_{[1, 1, 0]} - \chi_{[4, 0, 1]} + \chi_{[7, 0, 0]})\hat q^{\frac{7}{2}}
\nonumber\\&
+(-2 - \chi_{[1, 0, 1]} + 2\chi_{[2, 1, 0]} - \chi_{[5, 0, 1]} + \chi_{[8, 0, 0]})\hat q^4
+{\cal O}(\hat q^{\frac{9}{2}}).
\label{abjmn1}
\end{align}
\begin{align}
{\cal I}^{\rm grav}_{N=1}
&=(\cdots\mbox{terms identical to (\ref{abjmn1})}\cdots)
\nonumber\\&
+(\chi_{[0,0,4]}+\chi_{[0,2,0]}+\chi_{[0,4,0]}+\chi_{[0,6,0]}-\chi_{[1,0,1]}+\chi_{[2,0,2]}
+2 \chi_{[2,1,0]}
\nonumber\\&\qquad
+\chi_{[2,2,2]}+\chi_{[4,0,0]}
+\chi_{[4,2,0]}+\chi_{[4,4,0]}-\chi_{[5,0,1]}+\chi_{[6,0,2]}+2 \chi_{[8,0,0]}
\nonumber\\&\qquad
+\chi_{[8,2,0]}+\chi_{[12,0,0]}-1)\hat{q}^4
+{\cal O}(\hat q^{\frac{9}{2}}).
\end{align}
\begin{align}
{\cal I}^{\rm ABJM}_{N=2}
&=1
+\chi_{[1, 0, 0]}\hat q^{\frac{1}{2}}
+2\chi_{[2, 0, 0]}\hat q
+(-\chi_{[0, 0, 1]} + \chi_{[1, 1, 0]} + 2\chi_{[3, 0, 0]})\hat q^{\frac{3}{2}}
\nonumber\\&
+(-1 + \chi_{[0, 2, 0]} - 2\chi_{[1, 0, 1]} + \chi_{[2, 1, 0]} + 3\chi_{[4, 0, 0]})\hat q^2
\nonumber\\&
+(-\chi_{[0, 1, 1]} - 2\chi_{[1, 0, 0]} + \chi_{[1, 2, 0]} - 3\chi_{[2, 0, 1]} +
         2\chi_{[3, 1, 0]} + 3\chi_{[5, 0, 0]})\hat q^{\frac{5}{2}}
\nonumber\\&
+(\chi_{[0, 1, 0]} - 2\chi_{[1, 1, 1]} -
         3\chi_{[2, 0, 0]} + 2\chi_{[2, 2, 0]} - 4\chi_{[3, 0, 1]} + 2\chi_{[4, 1, 0]} +
         4\chi_{[6, 0, 0]})\hat q^3
\nonumber\\&
+(2\chi_{[0, 0, 1]} - \chi_{[0, 2, 1]} + \chi_{[1, 0, 2]} +
         2\chi_{[1, 1, 0]} + \chi_{[1, 3, 0]} - 3\chi_{[2, 1, 1]} - 4\chi_{[3, 0, 0]}
\nonumber\\&\qquad          +
         2\chi_{[3, 2, 0]} - 5\chi_{[4, 0, 1]} + 3\chi_{[5, 1, 0]} + 4\chi_{[7, 0, 0]})\hat q^{\frac{7}{2}}
\nonumber\\&
+(-2 + \chi_{[0, 1, 2]} + \chi_{[0, 2, 0]} + \chi_{[0, 4, 0]} + 5\chi_{[1, 0, 1]} -
         2\chi_{[1, 2, 1]} + \chi_{[2, 0, 2]} + 4\chi_{[2, 1, 0]}
\nonumber\\&\qquad
          + \chi_{[2, 3, 0]} -
         4\chi_{[3, 1, 1]} - 5\chi_{[4, 0, 0]} + 3\chi_{[4, 2, 0]} - 6\chi_{[5, 0, 1]} +
         3\chi_{[6, 1, 0]} + 5\chi_{[8, 0, 0]})\hat q^4
\nonumber\\&
+(\chi_{[0, 1, 1]} - \chi_{[0, 3, 1]} -
         4\chi_{[1, 0, 0]} + \chi_{[1, 1, 2]} + 3\chi_{[1, 2, 0]} + \chi_{[1, 4, 0]}
\nonumber\\&\qquad
         +
         7\chi_{[2, 0, 1]}
          - 3\chi_{[2, 2, 1]} + 2\chi_{[3, 0, 2]} + 5\chi_{[3, 1, 0]} +
         2\chi_{[3, 3, 0]} - 5\chi_{[4, 1, 1]}
\nonumber\\&\qquad
          - 6\chi_{[5, 0, 0]} + 3\chi_{[5, 2, 0]} -
         7\chi_{[6, 0, 1]} + 4\chi_{[7, 1, 0]} + 5\chi_{[9, 0, 0]})\hat q^{\frac{9}{2}}
\nonumber\\&
+(-2\chi_{[0, 0, 2]} - 6\chi_{[0, 1, 0]} + 2\chi_{[0, 3, 0]} - 2\chi_{[1, 3, 1]} -
         9\chi_{[2, 0, 0]} + 2\chi_{[2, 1, 2]}
\nonumber\\&\qquad
      + 4\chi_{[2, 2, 0]} + 2\chi_{[2, 4, 0]} +
         9\chi_{[3, 0, 1]} - 4\chi_{[3, 2, 1]} + 2\chi_{[4, 0, 2]} + 7\chi_{[4, 1, 0]}
\nonumber\\&\qquad
          +
         2\chi_{[4, 3, 0]} - 6\chi_{[5, 1, 1]} - 7\chi_{[6, 0, 0]} + 4\chi_{[6, 2, 0]} -
         8\chi_{[7, 0, 1]} + 4\chi_{[8, 1, 0]}
\nonumber\\&\qquad
          + 6\chi_{[10, 0, 0]})\hat q^5
+{\cal O}(\hat q^{\frac{11}{2}}).
\label{abjmn2}
\end{align}
\begin{align}
{\cal I}^{\rm grav}_{N=2}&=(\cdots\mbox{terms identical to (\ref{abjmn2})}\cdots)
\nonumber\\&
   +(-\chi_{[0,0,2]}-6 \chi_{[0,1,0]}+\chi_{[0,2,2]}+2 \chi_{[0,3,0]}+\chi_{[0,4,2]}-2 \chi_{[1,3,1]}-8 \chi_{[2,0,0]}
\nonumber\\&\qquad
   +\chi_{[2,0,4]}+2 \chi_{[2,1,2]}+5 \chi_{[2,2,0]}+3 \chi_{[2,4,0]}+\chi_{[2,6,0]}+9 \chi_{[3,0,1]}-4 \chi_{[3,2,1]}
\nonumber\\&\qquad
   +3 \chi_{[4,0,2]}+7 \chi_{[4,1,0]}
   +\chi_{[4,2,2]}+2 \chi_{[4,3,0]}-6 \chi_{[5,1,1]}-6 \chi_{[6,0,0]}+5 \chi_{[6,2,0]}
\nonumber\\&\qquad
   +\chi_{[6,4,0]}-8 \chi_{[7,0,1]}+\chi_{[8,0,2]}
   +4 \chi_{[8,1,0]}+7 \chi_{[10,0,0]}+\chi_{[10,2,0]}+\chi_{[14,0,0]})\hat{q}^5
\nonumber\\&
   +{\cal O}(\hat{q}^{\frac{11}{2}}).
\end{align}
\begin{align}
{\cal I}_{N=3}^{\rm ABJM}
&=
1+\chi_{[1,0,0]} \hat{q}^{\frac{1}{2}}+2 \chi_{[2,0,0]} q+(-\chi_{[0,0,1]}+\chi_{[1,1,0]}+3 \chi_{[3,0,0]}) \hat{q}^{\frac{3}{2}}
\nonumber\\&
+(2 \chi_{[0,2,0]}-2 \chi_{[1,0,1]}+2 \chi_{[2,1,0]}+4 \chi_{[4,0,0]}-1)\hat{q}^2
\nonumber\\&
+(-2 \chi_{[1,0,0]}+3 \chi_{[1,2,0]}-4 \chi_{[2,0,1]}+4 \chi_{[3,1,0]}+5 \chi_{[5,0,0]}) \hat{q}^{\frac{5}{2}}
\nonumber\\&
+(\chi_{[0,0,2]}+\chi_{[0,3,0]}-3 \chi_{[1,1,1]}-5 \chi_{[2,0,0]}+6 \chi_{[2,2,0]}-5 \chi_{[3,0,1]}+5 \chi_{[4,1,0]}+7 \chi_{[6,0,0]}) \hat{q}^3
\nonumber\\&
+(\chi_{[0,0,1]}-4 \chi_{[0,2,1]}+\chi_{[1,0,2]}-3 \chi_{[1,1,0]}+4 \chi_{[1,3,0]}-4 \chi_{[2,1,1]}-8 \chi_{[3,0,0]}+8 \chi_{[3,2,0]}
\nonumber\\&\qquad
-8 \chi_{[4,0,1]}+8 \chi_{[5,1,0]}+8 \chi_{[7,0,0]}) \hat{q}^{\frac{7}{2}}
\nonumber\\&
+(-4 \chi_{[0,2,0]}+4 \chi_{[0,4,0]}+4 \chi_{[1,0,1]}-6 \chi_{[1,2,1]}+2 \chi_{[2,0,2]}-4
   \chi_{[2,1,0]}+6 \chi_{[2,3,0]}
\nonumber\\&\qquad
   -8 \chi_{[3,1,1]}-13 \chi_{[4,0,0]}+12 \chi_{[4,2,0]}-10 \chi_{[5,0,1]}+10 \chi_{[6,1,0]}+10 \chi_{[8,0,0]}-2) \hat{q}^4
\nonumber\\&
   +(5 \chi_{[0,1,1]}-\chi_{[0,3,1]}-\chi_{[1,0,0]}+\chi_{[1,1,2]}-4 \chi_{[1,2,0]}+6 \chi_{[1,4,0]}+10 \chi_{[2,0,1]}-12 \chi_{[2,2,1]}
\nonumber\\&\qquad
   +3 \chi_{[3,0,2]}-8 \chi_{[3,1,0]}+11 \chi_{[3,3,0]}-11 \chi_{[4,1,1]}-18 \chi_{[5,0,0]}+15 \chi_{[5,2,0]}-13 \chi_{[6,0,1]}
\nonumber\\&\qquad
   +13 \chi_{[7,1,0]}+12 \chi_{[9,0,0]}) \hat{q}^{\frac{9}{2}}
\nonumber\\&
   +(\chi_{[0,0,2]}+4 \chi_{[0,1,0]}+2 \chi_{[0,2,2]}+2 \chi_{[0,3,0]}+2
   \chi_{[0,5,0]}+13 \chi_{[1,1,1]}-8 \chi_{[1,3,1]}
\nonumber\\&\qquad
   -2 \chi_{[2,0,0]}+2 \chi_{[2,1,2]}-8 \chi_{[2,2,0]}+12 \chi_{[2,4,0]}+18 \chi_{[3,0,1]}-16 \chi_{[3,2,1]}+4 \chi_{[4,0,2]}
\nonumber\\&\qquad
   -10 \chi_{[4,1,0]}+14 \chi_{[4,3,0]}-16 \chi_{[5,1,1]}-25 \chi_{[6,0,0]}+20 \chi_{[6,2,0]}-16 \chi_{[7,0,1]}
\nonumber\\&\qquad
   +16 \chi_{[8,1,0]}+14 \chi_{[10,0,0]}) \hat{q}^5
\nonumber\\&
   +(2 \chi_{[0,0,1]}+8 \chi_{[0,2,1]}-8
   \chi_{[0,4,1]}-\chi_{[1,0,2]}-3 \chi_{[1,1,0]}+3 \chi_{[1,2,2]}-\chi_{[1,3,0]}+8 \chi_{[1,5,0]}
\nonumber\\&\qquad
   +23 \chi_{[2,1,1]}-12 \chi_{[2,3,1]}-2 \chi_{[3,0,0]}+4 \chi_{[3,1,2]}-8 \chi_{[3,2,0]}+16 \chi_{[3,4,0]}+29 \chi_{[4,0,1]}
\nonumber\\&\qquad
   -24 \chi_{[4,2,1]}+5 \chi_{[5,0,2]}-15 \chi_{[5,1,0]}+20 \chi_{[5,3,0]}-20 \chi_{[6,1,1]}-32 \chi_{[7,0,0]}+24 \chi_{[7,2,0]}
\nonumber\\&\qquad
   -20 \chi_{[8,0,1]}+20 \chi_{[9,1,0]}+16 \chi_{[11,0,0]}) \hat{q}^{\frac{11}{2}}
\nonumber\\&
   +(-6 \chi_{[0,1,2]}-14 \chi_{[0,2,0]}+\chi_{[0,3,2]}-4 \chi_{[0,4,0]}+7 \chi_{[0,6,0]}-10 \chi_{[1,0,1]}+17 \chi_{[1,2,1]}
\nonumber\\&\qquad
   -13 \chi_{[1,4,1]}-4 \chi_{[2,0,2]}-11 \chi_{[2,1,0]}+6 \chi_{[2,2,2]}+2 \chi_{[2,3,0]}+12 \chi_{[2,5,0]}+36 \chi_{[3,1,1]}
\nonumber\\&\qquad
   -21 \chi_{[3,3,1]}-5 \chi_{[4,0,0]}+5 \chi_{[4,1,2]}-12 \chi_{[4,2,0]}+23 \chi_{[4,4,0]}+42 \chi_{[5,0,1]}-30 \chi_{[5,2,1]}
\nonumber\\&\qquad
   +7 \chi_{[6,0,2]}-18 \chi_{[6,1,0]}+25 \chi_{[6,3,0]}-27 \chi_{[7,1,1]}-41 \chi_{[8,0,0]}+30 \chi_{[8,2,0]}-23 \chi_{[9,0,1]}
\nonumber\\&\qquad
   +23 \chi_{[10,1,0]}+19 \chi_{[12,0,0]}) \hat{q}^6+{\cal O}(\hat{q}^{\frac{13}{2}}).
\label{abjmn3}
\end{align}
\begin{align}
{\cal I}^{\rm grav}_{N=3}
&=(\cdots\mbox{terms identical with (\ref{abjmn3})}\cdots)
\nonumber\\&
+(\chi_{[0,0,4]}-6 \chi_{[0,1,2]}-13 \chi_{[0,2,0]}+\chi_{[0,2,4]}+\chi_{[0,3,2]}-3 \chi_{[0,4,0]}+8 \chi_{[0,6,0]}+\chi_{[0,8,0]}
\nonumber\\&\qquad
-10 \chi_{[1,0,1]}+17 \chi_{[1,2,1]}-13 \chi_{[1,4,1]}-3 \chi_{[2,0,2]}-11 \chi_{[2,1,0]}+7 \chi_{[2,2,2]}+2 \chi_{[2,3,0]}
\nonumber\\&\qquad
+\chi_{[2,4,2]}+12 \chi_{[2,5,0]}+36 \chi_{[3,1,1]}-21 \chi_{[3,3,1]}-4 \chi_{[4,0,0]}+\chi_{[4,0,4]}+5 \chi_{[4,1,2]}
\nonumber\\&\qquad
-11 \chi_{[4,2,0]}+24 \chi_{[4,4,0]}+\chi_{[4,6,0]}+42 \chi_{[5,0,1]}-30 \chi_{[5,2,1]}+8 \chi_{[6,0,2]}-18 \chi_{[6,1,0]}
\nonumber\\&\qquad
+\chi_{[6,2,2]}+25 \chi_{[6,3,0]}-27 \chi_{[7,1,1]}-40 \chi_{[8,0,0]}+31 \chi_{[8,2,0]}+\chi_{[8,4,0]}-23 \chi_{[9,0,1]}
\nonumber\\&\qquad
+\chi_{[10,0,2]}+23 \chi_{[10,1,0]}+20 \chi_{[12,0,0]}+\chi_{[12,2,0]}+\chi_{[16,0,0]}+1) \hat{q}^6
+{\cal O}(\hat{q}^{\frac{13}{2}}).
\end{align}

\subsection{$k=2$}

In the case of $k=2$ the system still has ${\cal N}=8$ supersymmetry,
and the index can be expanded in terms of $su(4)$ characters.
\begin{align}
{\cal I}^{{\rm ABJM}(0/2)}_{N=1}
&=
 1 + \chi_{[2, 0, 0]} \hat q +  (-1 - \chi_{[1, 0, 1]} + \chi_{[4, 0, 0]})\hat q^2
+{\cal O}(\hat q^3).
\end{align}
\begin{align}
{\cal I}^{{\rm ABJM}(0/2)}_{N=2}
&=
 1 +  \chi_{[2, 0, 0]} \hat q+
  (\chi_{[0, 2, 0]} - \chi_{[1, 0, 1]} + 2 \chi_{[4, 0, 0]})\hat q^2  \nonumber\\
 &+ (-\chi_{[1, 1, 1]} - \chi_{[2, 0, 0]} + \chi_{[2, 2, 0]} - 2 \chi_{[3, 0, 1]} +
    \chi_{[4, 1, 0]} + 2 \chi_{[6, 0, 0]}) \hat q^3
+{\cal O}(\hat q^4).
\end{align}
\begin{align}
{\cal I}^{{\rm ABJM}(0/2)}_{N=3}
&=1
+\chi_{[2, 0, 0]}\hat q
+(\chi_{[0, 2, 0]} - \chi_{[1, 0, 1]} + 2\chi_{[4, 0, 0]})\hat q^2
\nonumber\\&
+(\chi_{[0, 0, 2]} - \chi_{[1, 1, 1]} + 2\chi_{[2, 2, 0]} - 2\chi_{[3, 0, 1]}
      + \chi_{[4, 1, 0]} + 3\chi_{[6, 0, 0]})\hat q^3
\nonumber\\&
+(-1 - \chi_{[0, 2, 0]} + 2\chi_{[0, 4, 0]} - 2\chi_{[1, 2, 1]} + \chi_{[2, 0, 2]}
    + \chi_{[2, 3, 0]} - 2\chi_{[3, 1, 1]}
\nonumber\\&\qquad
    - 2\chi_{[4, 0, 0]} + 4\chi_{[4, 2, 0]}
    - 4\chi_{[5, 0, 1]} + 2\chi_{[6, 1, 0]} + 4\chi_{[8, 0, 0]})\hat q^4
+{\cal O}(\hat q^{\frac{9}{2}}).
\end{align}
\begin{align}
{\cal I}_{\rm KK}^{\ZZ_2}&=
1+ \chi _{[2,0,0]}\hat q+ (\chi _{[0,2,0]}-\chi _{[1,0,1]}+2 \chi _{[4,0,0]})\hat q^2\nonumber \\
&+(\chi _{[0,0,2]}-\chi _{[1,1,1]}+2 \chi _{[2,2,0]}-2 \chi _{[3,0,1]}+\chi _{[4,1,0]}+3 \chi _{[6,0,0]})\hat q^3 \nonumber \\
&+ (3 \chi
   _{[0,4,0]}-2 \chi _{[1,2,1]}+2 \chi _{[2,0,2]}+\chi _{[2,3,0]}-2 \chi _{[3,1,1]}-\chi _{[4,0,0]}+5 \chi _{[4,2,0]}-4 \chi _{[5,0,1]}\nonumber \\
&\qquad +2 \chi _{[6,1,0]}+5 \chi _{[8,0,0]})\hat q^4
+{\cal O}(\hat q^5).
\end{align}
\begin{align}
{\cal I}^{{\rm ABJM}(1/2)}_{N=1}
&=\chi_{[1, 0, 0]}\hat q^{\frac{1}{2}}
+(-\chi_{[0, 0, 1]} + \chi_{[3, 0, 0]})\hat q^{\frac{3}{2}}
+(-\chi_{[2, 0, 1]} + \chi_{[5, 0, 0]})\hat q^{\frac{5}{2}}
\nonumber\\&
+(2\chi_{[1, 1, 0]} - \chi_{[4, 0, 1]} + \chi_{[7, 0, 0]})\hat q^{\frac{7}{2}}
\nonumber\\&
+(-\chi_{[0,1,1]}-3 \chi_{[1,0,0]}-\chi_{[2,0,1]}+2 \chi_{[3,1,0]}-\chi_{[6,0,1]}+\chi_{[9,0,0]})\hat{q}^{\frac{9}{2}}
+{\cal O}(\hat q^{\frac{11}{2}}).
\label{abjmk2n1b1}
\end{align}
\begin{align}
{\cal I}^{{\rm grav}(1/2)}_{N=1}
&=(\cdots\mbox{terms identical with (\ref{abjmk2n1b1})}\cdots)
\nonumber\\&
+(-\chi_{[0,1,1]}-\chi_{[0,3,1]}-3 \chi _{[1,0,0]}-\chi _{[2,0,1]}-\chi _{[3,0,2]}
\nonumber\\&
\qquad+2 \chi _{[3,1,0]}-\chi_{[3,3,0]}-\chi _{[5,2,0]}-\chi _{[6,0,1}])\hat{q}^{\frac{9}{2}}
+{\cal O}(\hat{q}^{\frac{11}{2}}).
\end{align}
\begin{align}
{\cal I}^{{\rm ABJM}(1/2)}_{N=2}
&=\chi_{[2, 0, 0]}\hat q
+(-\chi_{[1, 0, 1]} + \chi_{[2, 1, 0]} + \chi_{[4, 0, 0]})\hat q^2
\nonumber\\&
+(-\chi_{[1, 1, 1]} - \chi_{[2, 0, 0]} + \chi_{[2, 2, 0]} - 2\chi_{[3, 0, 1]}
    + \chi_{[4, 1, 0]} + 2\chi_{[6, 0, 0]})\hat q^3
\nonumber\\&
+(\chi_{[0, 2, 0]} + 2\chi_{[1, 0, 1]} - \chi_{[1, 2, 1]} + \chi_{[2, 0, 2]} + \chi_{[2, 1, 0]} + \chi_{[2, 3, 0]}
    - 2\chi_{[3, 1, 1]}
\nonumber\\&\qquad
- 2 \chi_{[4, 0, 0]} + \chi_{[4, 2, 0]} - 3\chi_{[5, 0, 1]} + 2 \chi_{[6, 1, 0]}
    + 2\chi_{[8, 0, 0]})\hat q^4
\nonumber\\&
+(-\chi_{[0,0,2]}-2 \chi_{[0,1,0]}+\chi_{[0,3,0]}+\chi_{[1,1,1]}-\chi_{[1,3,1]}
    -3 \chi_{[2,0,0]}+\chi_{[2,1,2]}
\nonumber\\&\qquad
    +2 \chi_{[2,2,0]}+\chi_{[2,4,0]}+4 \chi_{[3,0,1]}
    -2 \chi_{[3,2,1]}+\chi_{[4,0,2]}+3 \chi_{[4,1,0]}+\chi_{[4,3,0]}
\nonumber\\&\qquad
    -3 \chi_{[5,1,1]}
    -3 \chi_{[6,0,0]}+2 \chi_{[6,2,0]}-4 \chi_{[7,0,1]}+2 \chi_{[8,1,0]}+3 \chi_{[10,0,0]})\hat q^5
\nonumber\\&
+(-\chi_{[0,1,2]}-2 \chi_{[0,2,0]}+\chi_{[0,4,0]}-\chi_{[1,2,1]}-\chi_{[1,4,1]}
   -3 \chi_{[2,0,2]}-9 \chi_{[2,1,0]}
\nonumber\\&\qquad
   +\chi_{[2,2,2]}+2 \chi_{[2,3,0]}+\chi_{[2,5,0]}
   -2 \chi_{[3,3,1]}-7 \chi_{[4,0,0]}+\chi_{[4,1,2]}+4 \chi_{[4,2,0]}
\nonumber\\&\qquad
   +\chi_{[4,4,0]}
   +6 \chi_{[5,0,1]}-3 \chi_{[5,2,1]}+2 \chi_{[6,0,2]}+4 \chi_{[6,1,0]}+2 \chi_{[6,3,0]}
\nonumber\\&\qquad
   -4 \chi_{[7,1,1]}-4 \chi_{[8,0,0]}+2 \chi_{[8,2,0]}-5 \chi_{[9,0,1]}+3 \chi_{[10,1,0]}+3 \chi_{[12,0,0]}+1)\hat q^6
\nonumber\\&
+{\cal O}(\hat q^{\frac{13}{2}}).
\label{abjmk2n2b1}
\end{align}
\begin{align}
{\cal I}^{{\rm grav}(1/2)}_{N=2}
&=(\cdots\mbox{terms identical with (\ref{abjmk2n2b1})}\cdots)
\nonumber\\&
+(-\chi_{[0,0,4]}-\chi_{[0,1,2]}-2 \chi_{[0,2,0]}+\chi_{[0,4,0]}-\chi_{[0,6,0]}-2 \chi_{[1,2,1]}-\chi_{[1,4,1]}
\nonumber\\&\qquad
-3 \chi _{[2,0,2]}-9 \chi_{[2,1,0]}+2 \chi_{[2,3,0]}+\chi_{[2,5,0]}-3\chi_{[3,3,1}]-8 \chi_{[4,0,0]}+4 \chi_{[4,2,0]}
\nonumber\\&\qquad
+6 \chi_{[5,0,1]}-3 \chi_{[5,2,1]}+\chi_{[6,0,2]}+4 \chi_{[6,1,0]}+\chi _{[6,3,0]}-4 \chi_{[7,1,1]}
\nonumber\\&\qquad
-4 \chi_{[8,0,0]}+\chi_{[8,2,0]}-5\chi_{[9,0,1]}+3 \chi_{[10,1,0]}+2 \chi_{[12,0,0]}+1)\hat{q}^6
+{\cal O}(\hat{q}^{\frac{13}{2}}).
\end{align}
\begin{align}
{\cal I}^{{\rm ABJM}(1/2)}_{N=3}
&=
\chi _{[3,0,0]}\hat{q}^{\frac{3}{2}} +(\chi _{[1,2,0]}-\chi _{[2,0,1]}+\chi _{[3,1,0]}+\chi _{[5,0,0]})\hat{q}^{\frac{5}{2}} \nonumber \\
&+(-\chi _{[0,2,1]}-\chi _{[1,1,0]}+\chi _{[1,3,0]}-\chi _{[2,1,1]}-\chi _{[3,0,0]}\nonumber \\
&\qquad +2 \chi_{[3,2,0]}-2 \chi _{[4,0,1]}+2 \chi _{[5,1,0]}+2 \chi _{[7,0,0]})\hat{q}^{\frac{7}{2}} \nonumber \\
&+(\chi _{[0,1,1]}-\chi _{[1,2,0]}+\chi _{[1,4,0]}+\chi _{[2,0,1]}-3 \chi _{[2,2,1]}+\chi _{[3,0,2]}-2 \chi _{[3,1,0]}\nonumber \\
&\qquad +3
   \chi _{[3,3,0]}-3 \chi _{[4,1,1]}-3 \chi _{[5,0,0]}+4 \chi _{[5,2,0]}-3 \chi _{[6,0,1]}+3 \chi _{[7,1,0]}+3 \chi _{[9,0,0]})\hat{q}^{\frac{9}{2}} \nonumber \\
&+(\chi _{[0,0,1]}+2 \chi _{[0,2,1]}-2 \chi _{[0,4,1]}+\chi_{[1,0,2]}+\chi _{[1,1,0]}+\chi _{[1,2,2]}+2 \chi _{[1,5,0]}\nonumber \\
&\qquad +5 \chi _{[2,1,1]}-3 \chi _{[2,3,1]}+\chi _{[3,1,2]}-2 \chi _{[3,2,0]}+4 \chi _{[3,4,0]}+4 \chi _{[4,0,1]}\nonumber \\
&\qquad -6 \chi _{[4,2,1]}+\chi _{[5,0,2]}-4 \chi _{[5,1,0]}+5 \chi _{[5,3,0]}-5 \chi _{[6,1,1]}-6 \chi _{[7,0,0]}\nonumber \\
&\qquad +6 \chi _{[7,2,0]}-5 \chi _{[8,0,1]}+5 \chi _{[9,1,0]}+4 \chi _{[11,0,0]})\hat{q}^{\frac{11}{2}} \nonumber \\
&+(-\chi _{[0,0,3]}-5 \chi _{[0,1,1]}+2 \chi _{[0,3,1]}-\chi _{[0,5,1]}-2 \chi _{[1,0,0]}-2 \chi _{[1,1,2]}\nonumber \\
&\qquad -5 \chi _{[1,2,0]}+\chi _{[1,3,2]}+3 \chi _{[1,6,0]}-2 \chi _{[2,0,1]}+7 \chi _{[2,2,1]}-6 \chi _{[2,4,1]}\nonumber \\
&\qquad -\chi _{[3,1,0]}+2 \chi _{[3,2,2]}+5 \chi _{[3,5,0]}+12 \chi _{[4,1,1]}-7 \chi _{[4,3,1]}+2 \chi _{[5,1,2]}\nonumber \\
&\qquad -3 \chi _{[5,2,0]}+7 \chi _{[5,4,0]}+10 \chi _{[6,0,1]}-10 \chi _{[6,2,1]}+2 \chi _{[7,0,2]}\nonumber \\
&\qquad -6 \chi _{[7,1,0]}+8 \chi _{[7,3,0]}-8 \chi _{[8,1,1]}-10 \chi _{[9,0,0]}+9 \chi _{[9,2,0]}\nonumber \\
&\qquad -7 \chi _{[10,0,1]}+7 \chi _{[11,1,0]}+5 \chi _{[13,0,0]})\hat{q}^{\frac{13}{2}} \nonumber \\
&
+(2\chi_{[0,0,1]}+3 \chi_{[0,4,1]}-3 \chi_{[0,6,1]}+\chi_{[1,0,2]}+3 \chi_{[1,1,0]}-4 \chi_{[1,2,2]}-9 \chi_{[1,3,0]}
\nonumber\\&\qquad
+\chi_{[1,4,2]}+\chi_{[1,5,0]}+3 \chi_{[1,7,0]}-2 \chi_{[2,0,3]}-13
   \chi_{[2,1,1]}+8 \chi_{[2,3,1]}-6 \chi_{[2,5,1]}
\nonumber\\&\qquad
   -4 \chi_{[3,0,0]}-7 \chi_{[3,1,2]}-17 \chi_{[3,2,0]}+3 \chi_{[3,3,2]}+\chi_{[3,4,0]}+7 \chi_{[3,6,0]}-11 \chi_{[4,0,1]}
\nonumber\\&\qquad
   +14 \chi_{[4,2,1]}-12 \chi_{[4,4,1]}-3 \chi_{[5,0,2]}-10 \chi_{[5,1,0]}+4 \chi_{[5,2,2]}+\chi_{[5,3,0]}+9 \chi_{[5,5,0]}
\nonumber\\&\qquad
   +21 \chi_{[6,1,1]}-12 \chi_{[6,3,1]}-2 \chi_{[7,0,0]}+3 \chi_{[7,1,2]}-4 \chi_{[7,2,0]}+11 \chi_{[7,4,0]}
\nonumber\\&\qquad
   +18 \chi_{[8,0,1]}-15 \chi_{[8,2,1]}+3 \chi_{[9,0,2]}-9 \chi_{[9,1,0]}+12 \chi_{[9,3,0]}-12 \chi_{[10,1,1]}
\nonumber\\&\qquad
   -15 \chi_{[11,0,0]}+12 \chi_{[11,2,0]}-9 \chi_{[12,0,1]}+9 \chi_{[13,1,0]}+7 \chi_{[15,0,0]}) \hat{q}^{\frac{15}{2}}
   +{\cal O}(\hat{q}^{8}).
\label{abjmk2n3b1}
\end{align}
\begin{align}
{\cal I}^{{\rm grav}(1/2)}_{N=3}
&=
(\cdots\mbox{terms identical with (\ref{abjmk2n3b1})}\cdots)\nonumber\\
&+(2 \chi _{[0,0,1]}-\chi _{[0,3,3]}+3 \chi _{[0,4,1]}-3 \chi _{[0,6,1]}+3 \chi _{[1,1,0]}-4 \chi _{[1,2,2]}\nonumber \\
&\qquad -10 \chi _{[1,3,0]}+\chi _{[1,5,0]}+3 \chi _{[1,7,0]}-3 \chi _{[2,0,3]}-13 \chi _{[2,1,1]}+7\chi _{[2,3,1]}\nonumber \\
&\qquad -7 \chi _{[2,5,1]}-4 \chi _{[3,0,0]}-\chi _{[3,0,4]}-8 \chi _{[3,1,2]}-18 \chi _{[3,2,0]}+2 \chi _{[3,3,2]}\nonumber \\
&\qquad +\chi _{[3,4,0]}+6 \chi _{[3,6,0]}-11 \chi _{[4,0,1]}-\chi _{[4,1,3]}+13\chi _{[4,2,1]}-12 \chi _{[4,4,1]}\nonumber \\
&\qquad -3 \chi _{[5,0,2]}-11 \chi _{[5,1,0]}+2 \chi _{[5,2,2]}+\chi _{[5,3,0]}+8 \chi _{[5,5,0]}+21 \chi _{[6,1,1]}\nonumber \\
&\qquad -13 \chi _{[6,3,1]}-3 \chi _{[7,0,0]}+2 \chi _{[7,1,2]}-4 \chi _{[7,2,0]}+10 \chi _{[7,4,0]}+18 \chi _{[8,0,1]}\nonumber \\
&\qquad -15 \chi _{[8,2,1]}+2 \chi _{[9,0,2]}-9 \chi _{[9,1,0]}+11 \chi _{[9,3,0]}-12 \chi _{[10,1,1]}-15 \chi _{[11,0,0]}\nonumber \\
&\qquad +11 \chi _{[11,2,0]}-9 \chi _{[12,0,1]}+9 \chi _{[13,1,0]}+6 \chi _{[15,0,0]})\hat{q}^{\frac{15}{2}} +{\cal O}(\hat{q}^{\frac{17}{2}}).
\end{align}

\subsection{$k=3$}
If $k\geq 3$ the supersymmetry is ${\cal N}=6$.
Correspondingly, the R-symmetry is $so(6)=su(4)$, and
after the choice of the complex supercharge $\hat{\cal Q}$
the manifest symmetry becomes $so(2)\times so(4)=u(1)\times su(2)_1\times su(2)_2$.
Correspondingly, we define fugacities $u$, $u'$, and $u''$ for
$u(1)$, $su(2)_1$, and $su(2)_2$ by
\begin{align}
\hat u_1=uu',\qquad
\hat u_2=uu'^{-1},\qquad
\hat u_1=u^{-1}u'',\qquad
\hat u_2=u^{-1}u''^{-1}.
\end{align}
In the following we use the $so(4)$ characters $\chi_{a,b}\equiv\chi_a(u')\chi_b(u'')$.
\begin{align}
 {\cal I}^{{\rm ABJM}(0/3)}_{N=1} &=
 1 + \chi_{1, 1} \hat q
 + (u^{-3}\chi_{0, 3} + u^3\chi_{3, 0}) \hat q^\frac{3}{2} \nonumber\\
 &+ (-2 - \chi_{0, 2} - \chi_{2, 0} + \chi_{2, 2}) \hat q^2
+{\cal O}(\hat q^{\frac{5}{2}}).
\end{align}
\begin{align}
 {\cal I}^{{\rm ABJM}(0/3)}_{N=2} &=
 1 + \chi_{1, 1} \hat q
 + (u^{-3}\chi_{0, 3} + u^3\chi_{3, 0}) \hat q^\frac{3}{2}
 + (-\chi_{0, 2} - \chi_{2, 0} + 2 \chi_{2, 2}) \hat q^2 \nonumber\\
 &+ (2 u^{-3}\chi_{1, 4} + 2 u^3\chi_{4, 1}) \hat q^\frac{5}{2}  \nonumber\\
 &+ (u^{-6}(\chi_{0, 2} + 2 \chi_{0, 6}) - 2 \chi_{1, 1} - \chi_{1, 3} -
    \chi_{3, 1} + 3 \chi_{3, 3}  \nonumber\\
 &\qquad+ u^6(\chi_{2, 0} + 2 \chi_{6, 0})) \hat q^3
+{\cal O}(\hat q^{\frac{7}{2}}).
\end{align}
\begin{align}
{\cal I}^{{\rm ABJM}(0/3)}_{N=3}
&=1
+\chi_{1, 1}\hat q
+(u^3\chi_{3, 0}+u^{-3} \chi_{0, 3})\hat q^{\frac{3}{2}}
+(-\chi_{0, 2} - \chi_{2, 0} + 2 \chi_{2, 2})\hat q^2
\nonumber\\&
+2 (u^3\chi_{4, 1}+u^{-3}\chi_{1, 4})\hat q^{\frac{5}{2}}
\nonumber\\&
+ (u^6(\chi_{2, 0}+ 2\chi_{6, 0})
   - \chi_{1, 3} - \chi_{3, 1} + 4 \chi_{3, 3}
+u^{-6} (\chi_{0, 2}+ 2\chi_{0, 6})
)\hat q^3
\nonumber\\&
+(u^3(-\chi_{1, 0} + \chi_{1, 2} - \chi_{3, 0} - \chi_{5, 0} + 4 \chi_{5, 2})
\nonumber\\&\qquad
+u^{-3}(-\chi_{0, 1} - \chi_{0, 3} - \chi_{0, 5} + \chi_{2, 1} + 4 \chi_{2, 5}))
\hat q^{\frac{7}{2}}
\nonumber\\&
+ (u^6(\chi_{3, 1} + \chi_{5, 1} + 4\chi_{7, 1})
 -1 + \chi_{0, 4} - 3 \chi_{2, 2} - \chi_{2, 4}
 + \chi_{4, 0} - \chi_{4, 2}
\nonumber\\&\qquad
 + 7 \chi_{4, 4}
 + u^{-6}(\chi_{1, 1} + \chi_{1, 1} + \chi_{1, 3} + \chi_{1, 5} + 4 \chi_{1, 7}))
\hat q^4
+{\cal O}(\hat q^{\frac{9}{2}}).
\end{align}
\begin{align}
{\cal I}_{\rm KK}^{\ZZ_3}
&=1
+\chi_{1,1}\hat q
+(u^3\chi_{3,0}+u^{-3}\chi_{0,3})\hat q^{\frac{3}{2}}
+(-\chi_{0,2}-\chi_{2,0}+2\chi_{2,2})\hat q^2
\nonumber\\&
+2(u^3\chi_{4,1}+u^{-3}\chi_{1,4})\hat q^{\frac{5}{2}}
\nonumber\\&
+(u^6(\chi_{2,0}+2\chi_{6,0})-\chi_{1,3}-\chi_{3,1}+4\chi_{3,3}+u^{-6}(\chi_{0,2}+2\chi_{0,6}))\hat q^3
\nonumber\\&
+( u^3 (-\chi_{1,0}+\chi_{1,2}-\chi_{3,0}-\chi_{5,0}+4\chi_{5,2})
\nonumber\\&\qquad
    +u^{-3}(-\chi_{0,1}-\chi_{0,3}-\chi_{0,5}+\chi_{2,1}+4\chi_{2,5}))\hat q^{\frac{7}{2}}
\nonumber\\&
+(
+u^6(\chi_{1,1}+\chi_{3,1}+\chi_{5,1}+4\chi_{7,1})
+\chi_{0,4}-\chi_{2,2}-\chi_{2,4}+\chi_{4,0}
\nonumber\\&\qquad
-\chi_{4,2}+8\chi_{4,4}+1
+u^{-6}(\chi_{1,1}+\chi_{1,3}+\chi_{1,5}+4\chi_{1,7})
)\hat q^4
+{\cal O}(\hat q^{\frac{9}{2}}).
\end{align}
\begin{align}
{\cal I}^{{\rm ABJM}(1/3)}_{N=1}
&=u \chi_{1,0}\hat{q}^{\frac{1}{2}}+u^{-2}\chi_{0,2}\hat{q}
+u( \chi_{2,1}- \chi_{0,1})\hat{q}^{\frac{3}{2}}
\nonumber\\&
+\left(u^4 \chi_{4,0}-u^{-2}\chi_{1,1}+u^{-2}\chi_{1,3}\right)\hat{q}^2+\mathcal{O}(\hat{q}^{\frac{5}{2}}).
\label{abjmk3n1b1}
\end{align}
\begin{align}
{\cal I}^{{\rm grav}(1/3)}_{N=1}
&=(\cdots\mbox{terms identical with (\ref{abjmk3n1b1})}\cdots)
\nonumber\\&
+\left(u^{4} \chi_{4,0}-u^{-2}\chi_{1,1}+u^{-2}\chi_{1,3}-u^{-8}\chi_{0,2}-u^{-8}\chi_{0,6}\right)\hat{q}^2
+\mathcal{O}(\hat{q}^{\frac{5}{2}}).
\end{align}
\begin{align}
{\cal I}^{{\rm ABJM}(1/3)}_{N=2}
&= u^2 \chi_{2,0}\hat{q}+u^{-1}\chi_{1,2}\hat{q}^{\frac{3}{2}}
+\left(u^2 \chi_{3,1}+u^{-4}\chi_{0,4}+u^{-4}\right)\hat{q}^2
\nonumber\\&
+\left(u^5 \chi_{3,0}+u^5 \chi_{5,0}-u^{-1}\chi_{0,1}+2 u^{-1} \chi_{2,3}\right)\hat{q}^{\frac{5}{2}}
\nonumber\\&
+ \left(2u^{-4} \chi_{1,5}-u^2 \chi_{2,0}-u^2 \chi_{2,2}-u^2 \chi_{4,0}+3 u^2 \chi_{4,2}-u^2\right)\hat{q}^3
+\mathcal{O}(\hat{q}^{\frac{7}{2}}).
\label{abjmk3n2b1}
\end{align}
\begin{align}
{\cal I}^{{\rm grav}(1/3)}_{N=2}
&=(\cdots\mbox{terms identical with (\ref{abjmk3n2b1})}\cdots)
\nonumber\\&
+(2 u^{-4} \chi_{1,5}-u^2 \chi_{2,0}-u^2 \chi_{2,2}-u^2 \chi_{4,0}+3 u^2 \chi_{4,2}-u^2
\nonumber\\&
\qquad-u^{-10}\chi_{0,4}-u^{-10}\chi_{0,8}-u^{-10})\hat{q}^3
+\mathcal{O}(\hat{q}^{\frac{7}{2}}).
\end{align}
\begin{align}
{\cal I}^{{\rm ABJM}(1/3)}_{N=3}
&=u^3 \chi_{3,0}\hat{q}^{\frac{3}{2}} +\chi_{2,2}\hat{q}^2
+\left(u^3 \chi_{0,1}+u^3 \chi_{4,1}+u^{-3}\chi_{1,0}+u^{-3}\chi_{1,4}\right)\hat{q}^{\frac{5}{2}}
\nonumber\\&
+\left(u^6 \chi_{2,0}+u^6\chi_{4,0}+u^6 \chi_{6,0}+u^{-6}\chi_{0,2}+u^{-6}\chi_{0,6}+\chi_{1,3}+2 \chi_{3,3}\right)\hat{q}^3
\nonumber\\&
+\left(-u^3 \chi_{1,0}-u^3 \chi_{3,0}+u^3 \chi_{3,2}-u^3 \chi_{5,0}+3 u^3 \chi_{5,2}\right.
\nonumber\\&
\left.\qquad+u^{-3}\chi_{2,1}+u^{-3}\chi_{2,3}+3u^{-3} \chi_{2,5}\right)\hat{q}^{\frac{7}{2}}
\nonumber\\&
+\left(u^6 \chi_{3,1}+u^6 \chi_{5,1}+2 u^6\chi_{7,1}+2u^{-6} \chi_{1,3}+u^{-6}\chi_{1,5}+2u^{-6} \chi_{1,7}-2 \chi_{0,2}\right.
\nonumber\\&
\left.\qquad+\chi_{0,4}-\chi_{2,0}-3 \chi_{2,2}+2 \chi_{4,0}+5 \chi_{4,4}-1\right)\hat{q}^4
+\mathcal{O}(\hat{q}^\frac{9}{2}).
   \label{abjmk3n3b1}
\end{align}
\begin{align}
{\cal I}^{{\rm grav}(1/3)}_{N=3}
&=(\cdots\mbox{terms identical with (\ref{abjmk3n3b1})}\cdots)
\nonumber\\&
+\left(u^6 \chi_{3,1}+u^6 \chi_{5,1}+2 u^6 \chi_{7,1}+2u^{-6} \chi_{1,3}+u^{-6}\chi_{1,5}+2u^{-6} \chi_{1,7}-2\chi_{0,2}\right.
\nonumber\\&
\qquad+\chi_{0,4}-\chi_{2,0}-3 \chi_{2,2}+2 \chi_{4,0}+5 \chi_{4,4}-1
\nonumber\\&
\left.\qquad-u^{-12}\chi_{0,2}-u^{-12}\chi_{0,6}-u^{-12}\chi_{0,10}\right)\hat{q}^4
+\mathcal{O}(\hat{q}^\frac{9}{2}).
\end{align}

\section{Technical remarks on superconformal representations}\label{irrrep}
To calculate the index of superconformal representations we
mainly followed the procedure proposed in \cite{Cordova:2016emh}.
In the expansion in subsection \ref{6dresults.sec}
the D-type and B-type representations appear.
We used in the main text the notations in \cite{Beem:2015aoa}.
They correspond to those used in \cite{Cordova:2016emh} as follows.
\begin{align}
{\cal D}[a,b]=D_1[0,0,0]_{2a+2b}^{(b,a)},\quad
{\cal B}[a,b]_n=B_{\ell}[0,n,0]_{n+2a+2b+4}^{(b,a)}
\end{align}
where $\ell$ is the level of the primary null state.
It is $\ell=3$ for $n=0$ and
$\ell=1$ for $n\geq1$.

The Series of representations ${\cal D}[m,0]$ ($m=1,2,3,\ldots$) appear in the Kaluza-Klein spectrum in $AdS_7\times S^4$,
and have been well studied.
The superconformal index of each of them is%
\footnote{For ${\cal D}[1,0]$ and ${\cal D}[2,0]$ we use the definitions $\chi^{\check u}_{-1}=0$ and $\chi^{\check u}_{-2}=-1$.}
\begin{align}
{\cal D}[m,0]
&=\frac{\chi_m^{\check u}\check q^{2m}
-\chi_{m-1}^{\check u}\chi_{(0,1)}\check q^{2m+\frac{2}{3}}
+\chi^{\check u}_{m-2}\chi_{(1,0)}\check q^{2m+\frac{4}{3}}
-\chi^{\check u}_{m-3}\check q^{2m+2}
}
{(1-\check q^{\frac{4}{3}}\check y_1)(1-\check q^{\frac{4}{3}}\check y_2)(1-\check q^{\frac{4}{3}}\check y_3)}
\label{dnindex}
\end{align}
The index of the free tensor multiplet (\ref{freetensorindex}) is obtained by setting $m=1$,
and the single-particle index of Kaluza-Klein modes (\ref{kkspi}) is obtained by summing up (\ref{dnindex}) over $m\in\ZZ_{\geq1}$.
For $m=1$ (free tensor multiplet) and $m=2$ (stress tensor multiplet) some RS trial states
have negative coefficients.
They are interpreted as equations of motion and conservation laws.

Other D-type representations appearing in the expansion are
\begin{align}
{\cal D}[0,4]
&=\check q^8
-\chi^{\check u}_1\chi_{[0,1]}\check q^{\frac{26}{3}}
+(\chi_{[0,2]}+\chi_{[1,0]}+\chi^{\check u}_2\chi_{[1,0]})\check q^{\frac{28}{3}}
\nonumber\\&
-(\chi^{\check u}_1+\chi^{\check u}_3+2\chi^{\check u}_1\chi_{[1,1]})\check q^{10}
+{\cal O}(\check q^{\frac{32}{3}}),\\
{\cal D}[1,4]
&=\chi^{\check u}_1\check q^{10}+{\cal O}(\check q^{\frac{32}{3}}),\\
{\cal D}[3,2]
&=\chi^{\check u}_3\check q^{10}+{\cal O}(\check q^{\frac{32}{3}}).
\label{dseries}
\end{align}
For ${\cal D}[0,4]$ and ${\cal D}[1,4]$ the RS procedure works well,
and we obtain no RS trial weights with negative coefficients.
For ${\cal D}[3,2]$
we obtain many weights with negative coefficients.
In \cite{Cordova:2016emh} it is proposed that
such weights should be simply eliminated.
However, we found that this procedure gives $\check x$-dependent result.
Namely, the elimination spoils the Bose-Fermi degeneracy of states with $\check\Delta\neq0$.
Fortunately the elimination affects terms of order $\check q^{12}$ or higher,
and the lowest order of the $\check x$-dependent terms is $\check q^{\frac{38}{3}}$.
Therefore, we expect the term shown in
(\ref{dseries}) is correct.

The B-type representation appearing in the expansion is
\begin{align}
{\cal B}[2,0]_0
=\chi^{\check u}_2\chi_{[1,0]}\check q^{\frac{28}{3}}
-(\chi^{\check u}_1+\chi^{\check u}_1\chi_{[1,1]}+\chi^{\check u}_3\chi_{[1,1]})\check q^{10}
+{\cal O}(\check q^{\frac{32}{3}}).
\label{b20index}
\end{align}
For this representation we obtain many weights with negative coefficients.
We again found that the elimination of them causes the $\check x$-dependence of the result.
The elimination affects the terms of order $\check q^{10}$ or higher,
and the $\check x$-dependence appears at $\check q^{\frac{32}{3}}$.
(\ref{b20index}) is the index after the elimination.
Fortunately, terms shown in
(\ref{b20index}) do not depend on $\check x$.

We also calculated (\ref{b20index}) in another way.
For $n\geq1$ the primary null state of ${\cal B}[2,0]_n$ appears at level $\ell=1$,
and the procedure is much simpler than the case of $n=0$ for which the level of the
primary null state is $\ell=3$.
The RS procedure works well for such representations and
all generated weights have positive coefficients.
To obtain
${\cal B}[2,0]_0$
we simply substitute $n=0$ in the general formula for $n\geq 1$.
Although we have no justification for this ``continuation,''
this kind of continuation reproduces correct results in many cases.
Indeed, we obtained the result whose first few terms agree with (\ref{b20index}),
and this strongly suggests the correctness of (\ref{b20index}).


\end{document}